\documentclass[epj]{svjour}
\usepackage{graphics}
\usepackage{float}
\usepackage{hyperref}
\usepackage{graphicx}
\usepackage{dcolumn}
\usepackage{bm}
\usepackage{amsmath}
\usepackage{lineno}
\usepackage[dvipsnames]{xcolor}
\begin{document}
\title{Functional renormalization group study of the
quark-meson model with {$\omega$} and {$\rho$} vector mesons}
\author{Mohammed Osman\inst{1} \thanks{\email{mhdosmn183@mails.ccnu.edu.cn}} \and Defu Hou\inst{1}\thanks{\email{houdf@mail.ccnu.edu.cn} (corresponding author)} \and Wentao Wang\inst{1}\thanks{\email{wwt@mails.ccnu.edu.cn}} \and Hui Zhang\inst{2}\thanks{\email{Mr.zhanghui@m.scnu.edu.cn}}% etc
% \thanks{\emph{Present address:} Insert the address here if needed}%
}                     % Do not remove
%
% \offprints{}          % Insert a name or remove this line
%
% \email{mhdosmn183@mails.ccnu.edu.cn}

\institute{Institute of Particle Physics (IOPP) and key laboratory of Quark and lepton physics
(MOE), Central China Normal University, Wuhan 430079, China \and  Key Laboratory of Atomic and Subatomic Structure and Quantum Control (MOE),
Guangdong Basic Research Center of Excellence for Structure and Fundamental
Interactions of Matter, Institute of Quantum Matter, South China Normal
University, Guangzhou 510006, China\\
 Guangdong-Hong Kong Joint Laboratory of Quantum Matter, Guangdong
Provincial Key Laboratory of Nuclear Science, Southern Nuclear Science Computing
Center, South China Normal University, Guangzhou 510006, China \\
Physics Department and Center for Exploration of Energy and Matter, Indiana University, 2401 N Milo B. Sampson Lane, Bloomington, IN 47408, USA.}

\date{Received: date / Revised version: date}
\abstract{
We employ the functional renormalization group flow equations to investigate the phase structure of the two-flavor quark-meson model in the presence of a finite isospin chemical potential, incorporating interactions with omega and rho vector mesons.
For comparison, we also compute the phase diagram in the chiral limit using the mean-field approximation.
Our findings demonstrate that omega and rho mesons affect the phase structure in markedly distinct ways, and the introduction of an isospin chemical potential leads to significant modifications in the phase boundaries and critical region.
Increasing the isospin chemical potential lowers the tricritical points temperature, and tends to suppress the unphysical ``back-bending" of the FRG phase boundary at low temperature.
\PACS{
      {PACS-key}{discribing text of that key}   \and
      {PACS-key}{discribing text of that key}
     } % end of PACS codes
} %end of abstract
\authorrunning{Osman, Defu, Wang, and Zhang}
\titlerunning{FRG study of QM model with $\omega$ and $\rho$ vector mesons}
\maketitle
\section{INTRODUCTION}
\label{intro}
The phase diagram of Quantum Chromodynamics (QCD) has long been a subject of intensive theoretical and experimental investigation, offering insights into the understanding of strongly interacting matter under extreme conditions~\cite{Braun-Munzinger:2008szb,Rennecke:2015eba}. Since the pioneering conjecture by Cabibbo and Parisi in 1975~\cite{Cabibbo:1975ig}, our understanding of the QCD phase structure has advanced significantly.

While lattice QCD calculations and high-energy heavy-ion collisions have greatly clarified the nature of the QCD transition at high temperature and low baryon chemical potential, the structure of the phase diagram at high density remains less well understood. This limitation is primarily due to the fermion-sign problem, which obstructs straightforward lattice simulations at finite chemical potential~\cite{deForcrand:2009zkb}. To overcome this barrier, a range of theoretical approaches based on effective models have been developed to provide complementary insights.

Heavy-ion collision experiments conducted by the STAR Collaboration at RHIC~\cite{STAR:2011hyh,STAR:2014clz,STAR:2017tfy,STAR:2017sal} and the NA61/SHINE Collaboration at the CERN SPS~\cite{NA61SHINE:2021wba,Grebieszkow:2017gqx} have played a crucial role in probing the QCD phase structure. These experiments provide important constraints on the location of the phase boundary and the possible existence of a critical endpoint. Looking ahead, future facilities such as NICA~\cite{blaschke2016topical}, FAIR~\cite{CBM:2016kpk}, and J-PARC~\cite{SAKO20141158} are expected to extend our understanding by exploring regions of high baryon density, strong magnetic fields, and finite isospin asymmetry.

Effective QCD models, such as the Nambu-Jona-Lasinio (NJL) model and the quark-meson (QM) model, have proven invaluable in interpreting experimental results and exploring the QCD phase structure. However, traditional analyses based on the mean-field approximation often neglect essential fluctuation effects. 
Among several models extending beyond the conventional Mean Field approximation (MF)~\cite{Nikolov:1996jj,Nemoto:1999qf,Oertel:2000jp,Baacke:2003dk,Andersen:2008qk,Muller:2010am,Yamazaki:2012ux,Zacchi:2017ahv,CamaraPereira:2020ipu} is the functional renormalization group (FRG). The FRG method offers a nonperturbative approach to include both quantum and thermal fluctuations systematically~\cite{Nikolov:1996jj,Nemoto:1999qf,Oertel:2000jp,Baacke:2003dk,Andersen:2008qk,Muller:2010am,Yamazaki:2012ux,Zacchi:2017ahv,CamaraPereira:2020ipu,Dupuis:2020fhh}. This approach allows for a consistent scale evolution of effective interactions and has significantly improved the predictive power of effective QCD models.

Extensive studies have applied the FRG to chiral effective models, including both the NJL~\cite{Fukushima:2012xw,Braun:2011pp,Aoki:2014ola,Aoki:2015hsa} and QM models~\cite{Schaefer:2004en,Herbst:2013ail,Fu:2015naa,Herbst:2013ufa,Tripolt:2013jra,Jung:2016yxl,Andersen:2013swa}. Most of these investigations focus on two-flavor systems with scalar ($\sigma$) and pseudoscalar ($\pi$) fluctuations~\cite{Schaefer:2004en,Schaefer:2006sr,Herbst:2010rf}, while some have included isovector vector ($\rho$) and axial-vector ($a_{1}$) mesons~\cite{Eser:2015pka,Jung:2016yxl,Rennecke:2015eba}. However, the incorporation of $\omega$ and $\rho$ vector meson fluctuations within FRG frameworks has so far been limited, primarily explored in the context of Walecka-type nuclear models~\cite{Drews:2014wba,Drews:2014spa,Drews:2013hha}, and mostly at low temperature and density.

In the context of quark matter, mean-field analyses have demonstrated that the presence of $\omega$ and $\rho$ mesons can significantly modify the phase structure, affecting the location of the phase boundary and the critical endpoint~\cite{Fukushima:2008wg,Bratovic:2012qs,Lourenco:2012yv}. This motivates a thorough investigation into the stability of these results under the inclusion of fluctuations beyond the mean-field level.

In this work, we investigate the influence of $\omega$ and $\rho$ vector mesons on the chiral phase structure within a two-flavor QM model. Using the FRG formalism, we include fluctuations in the scalar, pseudoscalar, and vector channels to examine their effects on the phase boundary and the critical endpoint. In the chiral limit, we expect to observe a second-order transition at high temperature and low chemical potential, and a first-order transition at low temperature and high chemical potential. We also investigate the presence of a tricritical point (TCP), where the transition changes from second to first order~\cite{Schaefer:2004en,Herbst:2013ail,Lu:2015naa,Adhikari:2017ydi}.

This paper is organized as follows. Sec.~\ref{sec:2} introduces the two-flavor QM model with $\omega$ and $\rho$ meson interactions and outlines the MF and FRG formalisms. Sec.~\ref{sec:3} presents results and analyzes the effects of fluctuations and vector interactions on the chiral phase structure. Finally, Sec.~\ref{sec:4} summarizes our findings.

\section{THE QUARK-MESON MODEL WITH $\omega$ AND $\rho$ VECTOR MESONS}
\label{sec:2}
We consider the two-flavor quark-meson (QM) model with $\omega$ and $\rho$ vector mesons. The model is formulated in \\Minkowski space, with the Lagrangian given by

\begin{equation}\label{e2.0}
\begin{split}
\mathcal{L} & =\bar{\psi}\left(i \gamma_\mu \partial^\mu+\frac{\mu_I}{2} \gamma_0 \tau_3+\mu \gamma_0\right) \psi \\
& -\bar{\psi}\left[g_s\left(\sigma+i \gamma_5 \boldsymbol{\tau} \cdot \boldsymbol{\pi}\right)+\gamma_\mu\left(g_\omega \omega^\mu+g_\rho \boldsymbol{\tau} \cdot \boldsymbol{\rho}^\mu\right)\right] \psi \\
& +\frac{1}{2} \partial_\mu \sigma \partial^\mu \sigma+\frac{1}{2} \partial_\mu \boldsymbol{\pi} \partial^\mu \boldsymbol{\pi}-\frac{1}{4} F_{\mu \nu}^{(\omega)} F^{(\omega) \mu \nu}-\frac{1}{4} \boldsymbol{R}_{\mu \nu}^{(\rho)} \boldsymbol{R}^{(\rho) \mu \nu} \\
& -U(\sigma, \boldsymbol{\pi}, \omega_{\mu}, \boldsymbol{\rho}_{\mu}),
\end{split}
\end{equation}
where the quark field $\psi = (u, d)^{T}$ represents the two light flavors and couples to the isoscalar-scalar field $\sigma$ and the pseudoscalar pion field $\boldsymbol{\pi}$, which together transform as a four-component field $(\sigma, \boldsymbol{\pi})^T$ under the chiral group. Boldface notation denotes a vector, and $\boldsymbol{\tau} = (\tau_{1}, \tau_{2}, \tau_{3})$ are the Pauli matrices in isospin space.

The field strength tensors of the vector bosons $\omega_\mu$ and $\boldsymbol{\rho}_\mu$ are defined as:
$F^{(\omega)}_{\mu\nu}=\partial_\mu \omega_\nu-\partial_\nu \omega_\mu$ and $\boldsymbol{R}^{(\rho)}_{\mu\nu}=\partial_\mu \boldsymbol{\rho}_\nu-\partial_\nu \boldsymbol{\rho}_\mu-g_\rho \boldsymbol{\rho}_\mu \times \boldsymbol{\rho}_\nu$ respectively. 

% The quark field $\psi = (u, d)^T$ represents the two light flavors and couples to the isoscalar-scalar field $\sigma$ and the pseudoscalar pion field $\boldsymbol{\pi}$, which together transform as a four-component field $(\sigma, \boldsymbol{\pi})^T$ under the chiral group. Boldface notation denotes isospin triplet fields, and $\boldsymbol{\tau} = (\tau_{1}, \tau_{2}, \tau_{3})$ are the Pauli matrices in isospin space.

An isospin chemical potential is introduced via $\mu_{I} = \mu_{u} - \mu_{d}$. In the presence of background vector fields, only the temporal components $\omega_{0}$ and $\rho_{0}^{3}$ are non-vanishing. Under this configuration, the non-Abelian of $\boldsymbol{R}^{(\rho)}_{\mu\nu}$ does not contribute in practice. A Hubbard-Stratonovich transformation is employed to bosonize four-fermion interactions, leading to effective vector-isoscalar, $\omega_\mu$ and vector-isovector, $\boldsymbol{\rho}_{\mu}$ fields. 

The scalar and pseudoscalar meson fields $\boldsymbol{\pi}$ and $\sigma$ will be incorporated non-perturbatively, whilst $\omega_\mu$ and $\boldsymbol{\rho}_\mu$ will be treated as mean fields. 
These vector bosons conveniently parametrize unresolved short-distance dynamics. They are not to be identified with the physical omega and rho mesons.

The potential for $\sigma$, $\boldsymbol{\pi}$, $\omega_{\mu}$, and $\boldsymbol{\rho}_{\mu}$ is given by

\begin{equation}\label{e2.2}
\begin{split}
\displaystyle U(\sigma, \boldsymbol{\pi}, \omega_{\mu}, \boldsymbol{\rho}_{\mu})=&\frac{\lambda}{4}(\sigma^2+\boldsymbol{\pi}^2-f_{\pi}^2)^2\\
&-\frac{m^2_\omega}{2}\omega_\mu \omega^\mu-\frac{m^2_\rho}{2}\boldsymbol{\rho}_\mu \boldsymbol{\rho}^\mu,\\[5pt]
\end{split}
\end{equation}
where $f_{\pi} = 93$ MeV is the pion decay constant, and we use $m_{\omega} \approx m_{\rho} \sim 1$ GeV.\par
The relevant model parameters are $g_{s}$, $g_{\omega}$, $g_{\rho}$, and $\lambda$.
The values of these parameters may vary between the MF and FRG calculations when attempting to replicate the same value for quantities such as the constituent quark mass of $\sim 300$ MeV.  In our calculations, the values of $g_\omega$ and $g_\rho$ are consistently expressed as $g_{\omega}/m_{\omega}$ and $g_{\rho}/m_{\rho}$, respectively. Consequently, we will refrain from considering their values independently.
\subsection{Mean-field approximation}
In the vacuum, chiral symmetry is explicitly broken, and the expectation values of the meson fields are $\langle\sigma\rangle = f_{\pi}$ and $\langle\pi\rangle = 0$. Due to rotational symmetry, only the zero
component of the vector fields $\omega_{\mu}$ and $\boldsymbol{\rho}_{\mu}$ can have an expectation value~\cite{Floerchinger:2012xd}.
Restricting to these components ($\omega_{0}$ and $\rho^3_{0}$), the mean-field potential takes the form

\begin{equation}
\begin{array}{l}
\displaystyle U_{MF}(\sigma,\omega,\rho)=\frac{\lambda}{4}(\sigma^{2}-f^{2}_{\pi})^{2}-\frac{m^{2}_{\omega}}{2}(\omega_{0})^{2}-\frac{m^{2}_{\rho}}{2}(\rho_{0}^{3})^{2}.\\[5pt]
\end{array}
\label{e2.z}
\end{equation}

The mean-field effective potential is given by

\begin{equation}\label{e2.3}
\begin{array}{l}
\displaystyle \Omega_{MF}=\Omega_{\psi\bar{\psi}}+U_{MF}(\sigma,\omega_{0},\rho^{3}_{0}),
\end{array}
\end{equation}
with thermal quark and antiquark contributions. $\mu$ is the quark chemical potential, $T$ is the temperature, and $\beta=\frac{1}{T}$

\begin{equation}\label{e2.4}
\begin{split}
\Omega_{\psi \bar{\psi}}=- & \mathrm{\nu}_{\mathrm{q}} \int \frac{\mathrm{d}^{3} \mathrm{p}}{(2 \pi)^{3}}\left\{\mathrm{E}_{\mathrm{q}} \theta\left(\Lambda_{\mathrm{MF}}^{2}-\boldsymbol{p}^{2}\right)\right\} \\
& -\frac{\mathrm{\nu}_{\mathrm{q}}}{2} \mathrm{~T} \int \frac{d^{3} \mathrm{p}}{(2 \pi)^{3}}\left\{\ln \left[1+\mathrm{e}^{-\beta\left(\mathrm{E}_{\mathrm{q}}-\mu_{\mathrm{eff}}^{+}\right)}\right]\right. \\
& +\ln \left[1+\mathrm{e}^{-\beta\left(\mathrm{E}_{\mathrm{q}}+\mu_{\text{eff }}^{+ }\right)}\right] \\
& \left.+\ln \left[1+\mathrm{e}^{-\beta\left(\mathrm{E}_{\mathrm{q}}-\mu_{\mathrm{eff}}^{-}\right)}\right]+\ln \left[1+\mathrm{e}^{-\beta\left(\mathrm{E}_{\mathrm{q}}+\mu_{\mathrm{eff}}^{-}\right)}\right]\right\}.
\end{split}
\end{equation}

Here, $\nu_{q} = 2\,(\text{spin}) \times 2\,(\text{flavor}) \times 3\,(\text{color}) = 12$ is the quark degeneracy factor, and $E_{q} = \sqrt{p^{2} + m_{\text{eff}}^{2}}$ with effective quark mass $m_{\text{eff}} = g_{s}\sigma$. The first term is the fermion vacuum fluctuation; if we eliminated it, the chiral limit transition would always be of first order ~\cite{Skokov:2010sf}.

The effective chemical potentials for quarks and antiquarks are defined as:

\begin{equation}\label{e2.5}
\begin{array}{l}
\displaystyle \mu^{\pm}_{eff}=(\mu -g_\omega \omega_{0})\pm \Big(\frac{\mu_{I}}{2}+g_{\rho} \rho^{3}_{0}\Big),\\[5pt]
\end{array}
\end{equation}
For a given temperature $T$ and chemical potential $\mu$, the mean-field effective potential and the gap equation for $\omega_{0}$ and $\rho^{3}_{0}$ can be derived by solving the quantum equation of motion for $\omega_{0}$ and $\rho^{3}_{0}$:

\begin{equation}
\begin{split}
\omega_{0}=\frac{1}{2}\frac{g^{\omega}}{m^{2}_{\omega}}
v_{q}T\int\frac{d^{3}\mathbf{p}}{(2\pi)^{3}}
\Biggl[
\frac{\partial}{\partial\mu^{+}_{eff}}
\biggl(
\ln{[1+e^{-\beta(E_{q}-\mu_{eff}^{+})}]}\\
+\ln{[1+e^{-\beta(E_{q}+\mu_{eff}^{+})}]}
\biggr)\\
+\frac{\partial}{\partial\mu^{-}_{eff}}
\biggl(
\ln{[1+e^{-\beta(E_{q}-\mu_{eff}^{-})}]}
+\ln{[1+e^{-\beta(E_{q}+\mu_{eff}^{-})}]}
\biggr)
\Biggr],
\end{split}
\end{equation}

\begin{equation}\label{e2.12}
\begin{array}{l}
\displaystyle   \omega_{0}=\frac{1}{2}\frac{g_\omega}{m^2_\omega} [n^+ + n^-],
\end{array}
\end{equation}
and
\begin{equation}
\begin{split}
\label{e2.10}
\rho^{3}_{0}=\frac{1}{2}\frac{g^{\rho}}{m^{2}_{\rho}}
v_{q}T\int\frac{d^{3}\mathbf{p}}{(2\pi)^{3}}
\Biggl[
\frac{\partial}{\partial\mu^{+}_{eff}}
\biggl(
\ln{[1+e^{-\beta(E_{q}-\mu_{eff}^{-})}]}\\
+\ln{[1+e^{-\beta(E_{q}+\mu_{eff}^{-})}]}
\biggr)\\
-\frac{\partial}{\partial\mu^{-}_{eff}}
\biggl(
\ln{[1+e^{-\beta(E_{q}-\mu_{eff}^{+})}]}
+\ln{[1+e^{-\beta(E_{q}+\mu_{eff}^{+})}]}
\biggr)
\Biggr],
\end{split}
\end{equation}

\begin{equation}\label{e2.21}
\begin{array}{l}
\displaystyle \rho^{3}_{0}=\frac{1}{2}\frac{g_{\rho}}{m^2_{\rho}} [n^{-} - n^{+}].
\end{array}
\end{equation}
The up- and down-quark number densities $n^{+}$, $n^{-}$, respectively are computed from
\begin{equation}\label{e2.22}
\begin{array}{l}
\displaystyle  n^+(T,\mu_{eff}^+)=-\frac{\partial}{\partial\mu}\Omega_{\psi\overline{\psi}}(T,\mu_{eff}^+),
\end{array}
\end{equation}
and
\begin{equation}\label{e2.23}
\begin{array}{l}
\displaystyle  n^-(T,\mu_{eff}^-)=-\frac{\partial}{\partial\mu}\Omega_{\psi\overline{\psi}}(T,\mu_{eff}^-).
\end{array}
\end{equation}

In this framework, the vector couplings $g_{\omega}$ and $\rho^{3}_{0}$ appear only in the combinations $g_{\omega}/m_{\omega}$ and $g_{\rho}/m_{\rho}$. Their individual values are therefore not independently specified.
For numerical evaluation, we follow Ref.~\cite{Scavenius:2000qd} and adopt $g_{s} = 3.3$, $\lambda = 20$, which yield a constituent quark mass of $M_{\text{vac}} = g_{s}f_{\pi} \simeq 307$ MeV and a sigma mass of $m_{\sigma} = \sqrt{2\lambda f_{\pi}^{2}} \simeq 588$ MeV\par
In the mean-field approximation, $\omega_{0}$ and $\rho^{3}_{0}$ are directly proportional to the quark number density $n$. However, when fluctuations are included, the total quark density receives additional contributions beyond the single-particle level. As a result, this proportionality no longer holds.
The self-consistent equation for $\omega_{0}$ depends on the total quark density, while the equation for $\rho^{3}_{0}$ field depends on the difference in the quark densities in the MF. In symmetric matter ($\mu_{u} = \mu_{d}$), $\rho^{3}_{0}$ vanishes. This behavior is reproduced in both MF and FRG treatments.

While some works (e.g., Ref.~\cite{Drews:2014spa},) explored non-zero
UV values of $\rho^{3}_{0}$, introducing explicit isospin breaking, we follow a more conventional setup. We do not impose a nonzero ultraviolet value for $\rho^{3}_{0,\Lambda}$, consistent with the assumption of isospin symmetry in the high-energy limit. Similarly,
 although Ref. ~\cite{Zhang:2017icm} considered non-zero $\omega_{0,\Lambda}$, we focus
on dynamically generated vector fields in the IR regime.

\subsection{FRG flow equation}
The functional renormalization group (FRG) is a powerful non-perturbative method that allows the incorporation of quantum and thermal fluctuations in a field theory~\cite{Berges:2000ew}, and it has been extensively applied to effective QCD models~\cite{Schaefer:2004en,Schaefer:2006sr,Herbst:2010rf,tripolt2014spectral,Tripolt:2017zgc,Strodthoff:2013cua}. The effective average action $\Gamma_{k}$, which depends on the RG scale $k$, obeys the exact flow equation
\begin{equation}\label{e2.24}
\begin{array}{l}
\displaystyle  \partial_k \Gamma_k=\frac{1}{2}Tr\left[\frac{\partial_k R_k}{\Gamma^{(2)}_k+R_k}\right],
\end{array}
\end{equation}
where $\Gamma^{(2)}_{k}$ is the second functional derivative with respect to the fields, and the trace includes momentum integration and sums over internal indices. The infrared regulator $R_{k}$ is introduced to suppress fluctuations at momenta below the scale $k$.
While its form is arbitrary, it must satisfy constraints that ensure interpolation between the bare and full quantum effective actions.
In our setup, quarks, $\sigma$, and $\boldsymbol{\pi}$ serve as dynamical fields. The vector fields $\omega_{0}$ and $\rho^{3}_{0}$, in contrast, are non-dynamical because their temporal components are not coupled to time derivatives.
Therefore, the value of $\omega_{0}$ and $\rho_{0}^{3}$  are completely fixed by specifying the values of other fields. At each scale $k$, by solving consistency equations for given values of $\sigma$ and $\boldsymbol{\pi}$. As a result, $\omega_{0}$ and $\rho^{3}_{0}$ may be written as  $\sigma$, $\boldsymbol{\pi}$, and $k$, and appear in the effective chemical potential for quarks, affecting the dynamical fluctuations in the flow equations. Throughout this study, we neglect the flow of all wavefunction renormalization factors.

The scale-dependent effective potential is introduced by replacing the static potential $U$ with the running one $U_{k}$ in the Euclidean Lagrangian:
\begin{equation}\label{e2.25}
\Gamma_{k} = \int d^{4} x \, \mathcal{L} \big|_{{U \rightarrow U_{k}}}.
\end{equation}
Finite temperatures and chemical potentials are treated within the Matsubara formalism. The time coordinate is Wick-rotated as $t \rightarrow -i\tau$, and imaginary time is compactified with $\beta = 1/T$. Due to chiral symmetry, the potential depends on the $\sigma$ and $\pi$ only through the invariant:
\begin{equation}\label{e2.27}
\phi^{2} = \sigma^{2} + \boldsymbol{\pi}^{2}.
\end{equation}
As mentioned, the vector fields $\omega_{0}, \rho^{3}_{0}$ appear here only as mean fields. The complete $k$-dependence is in the effective potential  $U_{k}$. In analogy to the mean-field potential, the effective potential has a chirally symmetric piece, $U_{k}^{\phi}$ the explicit chiral symmetry breaking term and the mass terms of the vector bosons

\begin{equation}\label{e2.28}
U_{k} = U_{k}^{\phi} + U_{k}^{\omega} + U_{k}^{\rho}.
\end{equation}

Starting with some ultraviolet $(UV)$ potentials $U_{\Lambda}$ as our initial conditions, we integrate fluctuations and obtain the scale-dependent $U_{k}$,
The form of $U_{k}^{\phi}$ will be determined without assuming any specific forms, while for the potential of the $\omega$ and $\rho$ fields, we use the same form as in Eq.~\eqref{e2.z}:
\begin{equation}
U_{k}^{\omega,\rho} = -\frac{m_{\omega}^2}{2} \omega_{0,k}^{2} - \frac{m_{\rho}^{2}}{2} (\rho^{3}_{0,k})^{2}.
\end{equation}

These background fields are fixed at each RG scale via
\begin{equation}
\frac{\partial U_{k}}{\partial \omega_{0,k}} = 0, \qquad \frac{\partial U_{k}}{\partial \rho^{3}_{0,k}} = 0.
\end{equation}

To use Wetterich’s equation, a regulator function must be chosen that respects the interpolation limits of the effective average action. We employ the so-called optimized or Litim regulator function~\cite{Litim:2001up}, for bosons and fermions, respectively, given by:

\begin{align}
R_{k}^{B}(p) &= (k^{2} - \boldsymbol{p}^{2}) \, \theta(k^{2} - \boldsymbol{p}^{2}), \\
R_{k}^{F}(p) &= \left(
\begin{array}{cc}
0 & i p_{i} (\gamma^{E}_{i})^{T} \\
i p_{i} \gamma^{E}_{i} & 0
\end{array}
\right) \left(\sqrt{\frac{k^{2}}{p^{2}} - 1} \right) \theta(k^{2} - \boldsymbol{p}^{2}).
\end{align}
The flow equation for the potential becomes:
% \begin{align}
% \partial_{k} U_{k}^\phi(T,\mu) = \frac{k^{4}}{12\pi^{2}} \Bigg[
% &\frac{3(1 + 2 n_{B}(E_{\p}i))}{E_{\pi}} + \frac{1 + 2 n_{B}(E_{\sigma})}{E_{\sigma}}
% - v_{q} \left( \frac{1 - n_{F}(E_{q}, \mu^{+}_{\text{eff}}) - n_{F}(E_{q}, -\mu^{+}_{\text{eff}})}{E_{q}}
% + \frac{1 - n_{F}(E_{q}, \mu^{-}_{\text{eff}}) - n_{F}(E_{q}, -\mu^{-}_{\text{eff}})}{E_{q}} \right) \Bigg],
% \end{align}
\begin{equation}
\label{e2.32}
\begin{split}
\partial_{k}U_{k}^{\phi}(T,\mu)=\frac{k^{4}}{12\pi^{2}} 
\Biggl\{
\frac{3[1+2n_{B}(E_{\pi})]}{E_{q}}+\frac{[1+2n_{B}(E_{\sigma})]}{E_{\sigma}}\\
-v_{q}
\biggl[
\frac{1-n_{F}(E_{q},\mu^{+}_{eff})-n_{F}(E_{q},-\mu^{-}_{eff})}{E_{q}}
\biggr]\\
-v_{q}
\biggl[
\frac{1-n_{F}(E_{q},\mu^{+}_{eff})-n_{F}(E_{q},\mu^{-}_{eff})}{E_{q}}
\biggr]
\Biggr\},
\end{split}
\end{equation}
where the effective energies are given by:
\begin{equation}\label{e2.33}
\begin{array}{l}
\displaystyle   E_\pi=\sqrt{k^2+M^2_\pi},\\
\end{array}
\end{equation}

\begin{equation}\label{e2.34}
\begin{array}{l}
\displaystyle   E_\sigma=\sqrt{k^2+M^2_\sigma},\\
\end{array}
\end{equation}
  \begin{equation}\label{e2.35}
\begin{array}{l}
\displaystyle   E_q=\sqrt{k^2+M^2_q},\\
\end{array}
\end{equation}
for pion, sigma-meson, and quark, respectively. the scale-dependent particle masses are:

\begin{equation}\label{e2.36}
\begin{array}{l}
\displaystyle   M^2_q=g^2\phi^2,\\
\end{array}
\end{equation}

\begin{equation}\label{e2.37}
\begin{array}{cc}
\displaystyle   M^2_\pi=2U'_k(\phi^2),\\      
\end{array}
\end{equation}

\begin{equation}\label{e2.38}
\begin{array}{cc}
\displaystyle   M^2_\sigma=2U'_k(\phi^2)+4 \phi^2 U''_k(\phi^2),\\        
\end{array}
\end{equation}
while the effective quark chemical potentials are:

\begin{equation}\label{e2.6}
\begin{array}{l}
\displaystyle \mu^\pm_{eff}=(\mu -g_\omega \omega_{0,k})\pm (\frac{\mu_{I}
}{2}+g_\rho \rho^3_{0,k}),\\[5pt] 
\end{array}
\end{equation}
depend on the $k$ through $\omega_{0,k}$ and $\rho_{0,k}$. The boson and fermion occupation numbers are
\begin{equation}\label{e2.39}
\begin{array}{l}
\displaystyle   n_B(E)=\frac{1}{e^{\beta E}-1}, \ \ n_F(E,\mu)=\frac{1}{e^{\beta(E-\mu)}+1}.
\end{array}
\end{equation}
It may appear that the flow equation should be resolved jointly in the $\phi$, $\omega_0$, and $\rho^{3}_{0}$ directions. However, the $\omega_0$ and $\rho^{3}_{0}$ fields are non-dynamical, therefore, their flow equations can be determined for a specific value of $\phi$. This is similar to the Gauss law constraint in gauge theories. We determine for $\omega_{0,k}$ and $\rho^{3}_{0,k}$ at each momentum scale $k$~\cite{Drews:2014spa,Zhang:2017icm}:
\begin{equation}\label{e2.41}
\begin{array}{l}
\displaystyle \frac{\partial U_k}{\partial \omega_{0,k}}=0,\\
\displaystyle \frac{\partial U_k}{\partial \rho^3_{0,k}}=0.
\end{array}
\end{equation}

The flow equations become:
% \begin{align}
% \partial_{k} \rho^{3}_{0,k} &= -\frac{g_{\rho} k^{4}}{\pi^{2} m_{\rho}^{2} E_{q}} \left[
% -\frac{\partial}{\partial \mu^{+}_{\text{eff}}} \left( n_{F}(E_{q}, \mu^{+}_{\text{eff}}) + n_{F}(E_{q}, -\mu^{+}_{\text{eff}}) \right)
% + \frac{\partial}{\partial \mu^{-}_{\text{eff}}} \left( n_{F}(E_{q}, \mu^{-}_{\text{eff}}) + n_{F}(E_{q}, -\mu^{-}_{\text{eff}}) \right)
% \right], \\
% \partial_{k} \omega_{0,k} &= -\frac{g_{\omega} k^{4}}{\pi^{2} m_{\omega}^{2} E_{q}} \left[
% \frac{\partial}{\partial \mu^{+}_{\text{eff}}} \left( n_{F}(E_{q}, \mu^{+}_{\text{eff}}) + n_{F}(E_[q], -\mu^{+}_{\text{eff}}) \right)
% + \frac{\partial}{\partial \mu^{-}_{\text{eff}}} \left( n_{F}(E_{q}, \mu^{-}_{\text{eff}}) + n_{F}(E_{q}, -\mu^{-}_{\text{eff}}) \right)
% \right].
% \end{align}
\begin{equation}
\label{e2.46}
\begin{split}
\partial_{k}\rho^{3}_{0,k}=&-\frac{g_{\rho}k^{4}}{\pi^{2}m^{2}_{\rho}E_{q}}
\Biggl\{\\
&-\frac{\partial}{\partial\mu^{+}_{eff}}
\biggl(
n_{F}(E_{q},\mu^{+}_{eff})
+n_{F}(E_{q},-\mu^{+}_{eff})
\biggr)\\
&+\frac{\partial}{\partial\mu^{-}_{eff}}
\biggl(
n_{F}(E_{q},\mu^{-}_{eff})
+n_{F}(E_{q},-\mu^{-}_{eff})
\biggr)
\Biggr\},
\end{split}
\end{equation}

\begin{equation}
\label{e2.47}
\begin{split}
\partial_{k}\omega^{3}_{0,k}=&-\frac{g_{\omega}k^{4}}{\pi^{2}m^{2}_{\omega}E_{q}}
\Biggl\{\\
&\frac{\partial}{\partial\mu^{+}_{eff}}
\biggl(
n_{F}(E_{q},\mu^{+}_{eff})
+n_{F}(E_{q},-\mu^{+}_{eff})
\biggr)\\
&+\frac{\partial}{\partial\mu^{-}_{eff}}\biggl(
n_{F}(E_{q},\mu^{-}_{eff})
+n_{F}(E_{q},-\mu^{-}_{eff})
\biggr)
\Biggr\}.
\end{split}
\end{equation}
Note that the flow equations for $\omega_{0}$ and $\rho^{3}_{0}$ can be solved for a given $\phi$, independently of the potential $U^{k}_{\phi}$ (which only tells us where the minimum of $\phi$ is).\par
If the $k$-dependence of $\mu_{\text{eff}}$ is ignored, one recovers expressions similar to the MF densities.
But unlike the MF case, $\omega_{0}$ and $\rho^{3}_{0}$ are not directly proportional to the physical number density because the baryon density gets contributions not only from single particles but also fluctuations (see Eq.(~\ref{y})).
Moreover, as we will see in Sec.~\ref{sec:3} if we include the $k$ dependence in $\mu_{eff}$, the $\omega_{0,k}$ and $\rho^3_{0,k}$ fields at $k_{IR}$ are not even proportional to the single-particle contribution. Therefore, the extrapolation of the MF relation $\omega_{0}\sim {n}$ and $\rho_{0}\sim {n}$ does not work without exception to understand the FRG results.\par

The quark coupling constant $g_{s} = 3.2$. Differing from Ref.~\cite{Drews:2014spa}, given the qualitative nature of our work, the medium's impact on parameter selection is neglected. We established parameter values within a plausible range; these are not physically mandated but are selected phenomenologically to get the qualitative behavior. The typical range of $g_{\omega,\rho}/m_{\omega,\rho} \simeq 10^{-3} - 10^{-2} MeV^{-1}$ is the natural choice. We take it as an approximate optimal choice with a moderate value $6\times10^{-3} MeV^{-1}$ to show the influence of vector mesons on the phase diagram and values within that on the order parameters' results. We have checked that the overall qualitative result does not change the conclusion if we change the vector coupling constant to other values in that range.

Finally, the initial conditions for the flow equations must be established. The UV scale $\Lambda$ should be sufficiently large in order to take into account the relevant fluctuation effects and small enough to render the description in terms of the model degrees of freedom realistic~\cite{Drews:2013hha}. In our calculation we follow the choice of Ref.~\cite{Schaefer:2004en}, $\Lambda$ = 500 MeV. The initial for the potential is

\begin{equation}\label{x}
\begin{array}{l}
\displaystyle U^{\phi}_{\Lambda} = \frac{\lambda}{4}{\phi^4},
\end{array}
\end{equation}

 We calibrate the model by requiring that the vacuum of the effective potential $\sigma_{\text{vac}} \simeq f_{\pi}$ at zero temperature and zero chemical potential reproduces the physical pion decay constant, $f_\pi = 93\,\mathrm{MeV}$. This procedure leads to the determination of the initial condition for the quartic coupling in the chiral limit as $\lambda = 8$. Starting with the condition Eq.(~\ref{x}), the scale evolution first generates the $\phi^{2}$ terms, reflecting the universality.

The initial condition for the $\omega$ and $\rho$ fields has not been examined in detail, and we simply try.

\begin{equation}
    \begin{array}{cc}
         \displaystyle \omega_{0,\Lambda}(\phi) = 0, \qquad\displaystyle \rho^3_{0,\Lambda}(\phi) = 0,
    \end{array}
\end{equation}

Later, we will also present the result of another different initial condition, but it will turn out that such a modification does not change the main story in this paper.

Assembling all these elements, we calculate the effective potential with the fluctuations integrated to $k_{IR}$ = 0. The final step is to find $\Phi$ = $\sigma^{*}$, which minimizes the effective potential. At the minimum, the effective potential is identified as the thermodynamic potential,
\begin{equation}
     \begin{array}{l}
          \displaystyle \frac{T\Omega(\mu,T)}{V} = \Gamma_{IR=0}(\mu,T,\sigma^{*}).     
     \end{array}
 \end{equation}

In practice, it is numerically expensive to reduce the $IR$ cutoff and we typically stop the integration under
$k_{IR} \simeq20$ MeV.\par
The baryon number density is then obtained by taking the derivative with respect to $\mu_{B}$ = $N_{c}\mu$,
\begin{equation}\label{y}
     \begin{array}{l}
    \displaystyle n_{B}(\mu,T) = -\frac{1}{N}\frac{\partial\Omega_{{k}_{IR=0}}(\mu,T,\sigma^{*})}{\partial\mu},
     \end{array}
 \end{equation}
The derivative is taken numerically with the interval $\Delta\mu$ = 0.1 MeV.

\section{RESULTS}
\label{sec:3}
\subsection{Effective potential}
By numerically solving Eqs.~(\ref{e2.32}), (\ref{e2.46}), and (\ref{e2.47}), the results that can be obtained directly are the effective potentials at different energy scales. The evolution of the energy scale ranges from the set 500 MeV to the infrared cutoff less than or equal to 10 MeV. The evolution after the infrared cutoff is considered to remain unchanged within the accuracy.

\begin{figure}[!tbh]
\centering
\includegraphics[width=0.45\textwidth]{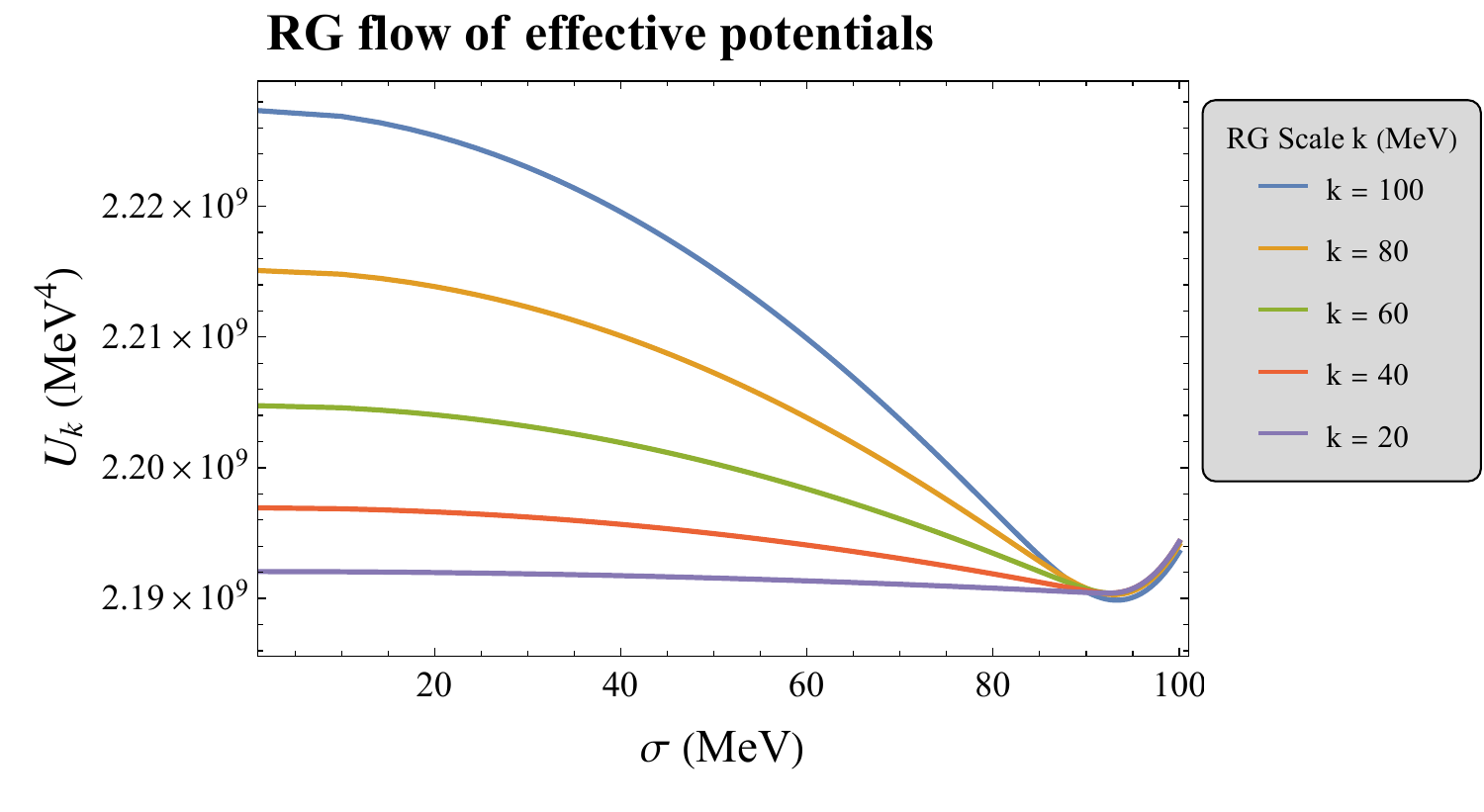}
\caption{\label{fig:1}The evolution of the effective potential with energy scales at $T = 20$ MeV, $\mu = 250$ MeV and $\mu_{I} = 0$ MeV.}
\end{figure}

\begin{figure}[!tbh]
\centering
\includegraphics[width=0.45\textwidth]{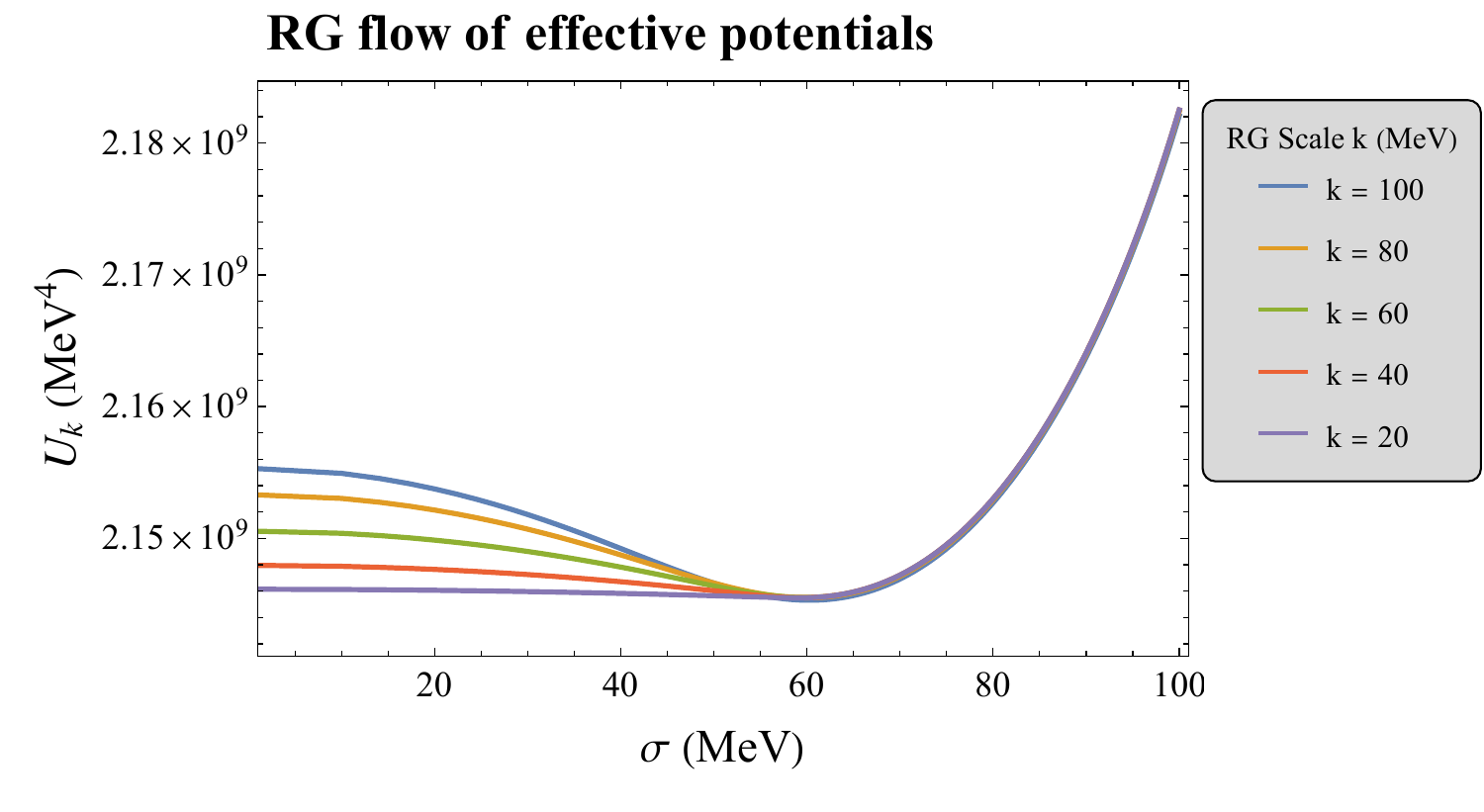}
 \caption{\label{fig:2}The evolution of the effective potential with energy scales at $T = 20$ MeV, $\mu = 250$ MeV and $\mu_{I} = 200$ MeV}

\end{figure}

To directly illustrate the impact of the isospin chemical potential $\mu_I$ on the flow of the effective potential $U_k$ and the formation of the physical vacuum, Fig.~\ref{fig:1} and Fig.~\ref{fig:2} present the evolution of $U_k$ as a function of the chiral condensate $\sigma$ with the momentum scale k. The introduction of $\mu_I$ serves two primary roles: on one hand, it reduces the magnitude of $\sigma$ the extreme value around $\phi=f_{\pi}$; on the other hand, it enhances the quantum and thermal fluctuations incorporated during the evolution of the renormalization group (RG) flow, thereby accelerating the approach of the flow toward the infrared fixed point. If one defines the RG time as $t_{RG}=ln\frac{\Lambda}{k}$, the system with a finite $\mu_I$ reaches the infrared cutoff at an ``earlier time'' of the flow.\par

\begin{figure}[!tbh]
\centering
\includegraphics[width=0.45\textwidth]{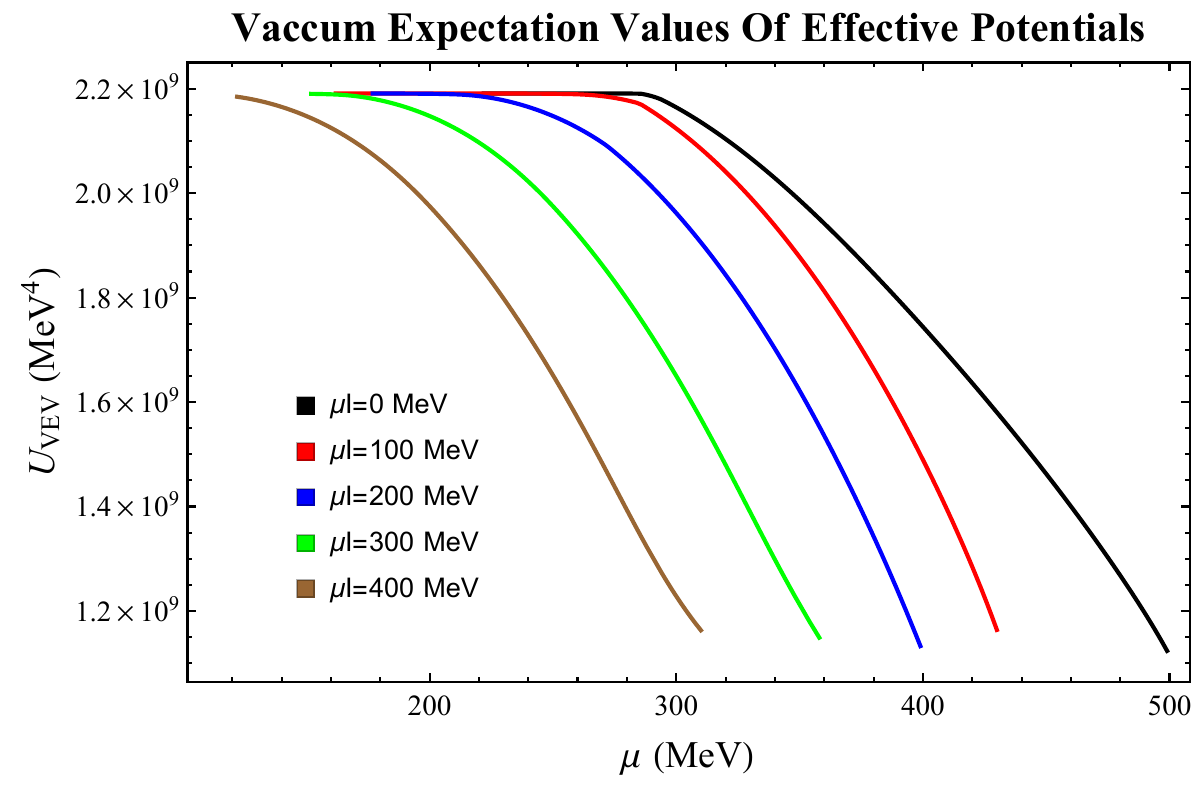}
\caption{\label{fig:3}The vacuum expectation value of the effective potential under different isospin chemical potentials, calculated at $T=10$ MeV, $g_{\omega}/m_{\omega}=g_{\rho}/m_{\rho}=0.006~[\mathrm{MeV}]^{-1}$. Different colored lines correspond to different isospin chemical potentials.}
\end{figure}

Fig.~\ref{fig:3} shows the relationship between the vacuum expectation values of corresponding effective potentials, $U_{VEV}$, defined by the global minimum of $U_k$ at IR cutoff, and the quark chemical potentials under different isospin chemical potentials at low temperatures. As $\mu_I$ increases, the quark chemical potential $\mu$ at which the effective potential exhibits a phase transition at a given temperature is progressively reduced.\\
\subsection{Chiral condensates}
\begin{figure}[!tbh]
\centering
\includegraphics[width=0.45\textwidth]{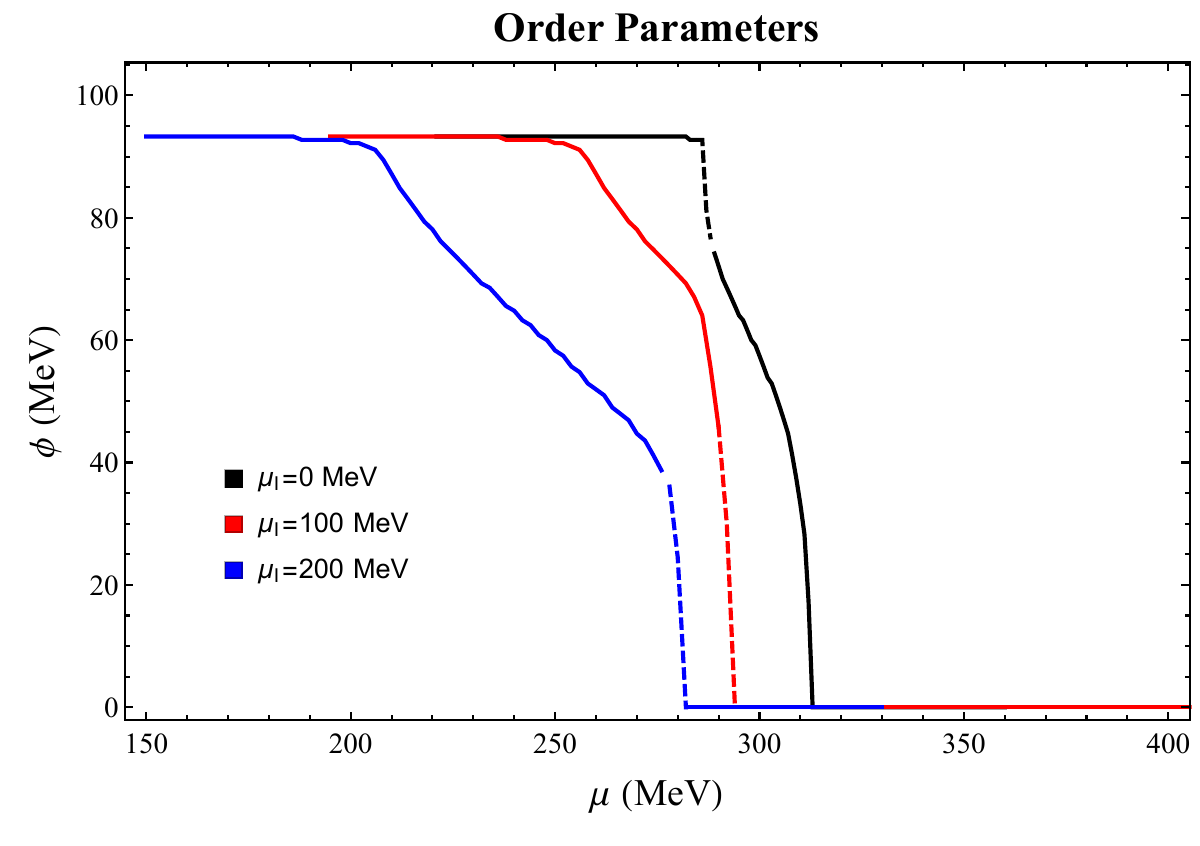}
\caption{\label{fig:4} Chiral condensates as a function of quark chemical potential under different isospin chemical potentials, calculated at $T=7$ MeV, $g_{\omega}/m_{\omega}=g_{\rho}/m_{\rho}=0.006~[\mathrm{MeV}]^{-1}$. Different colored lines correspond to different isospin chemical potentials.}
\end{figure}

\begin{figure}[!tbh]
\centering
\includegraphics[width=0.45\textwidth]{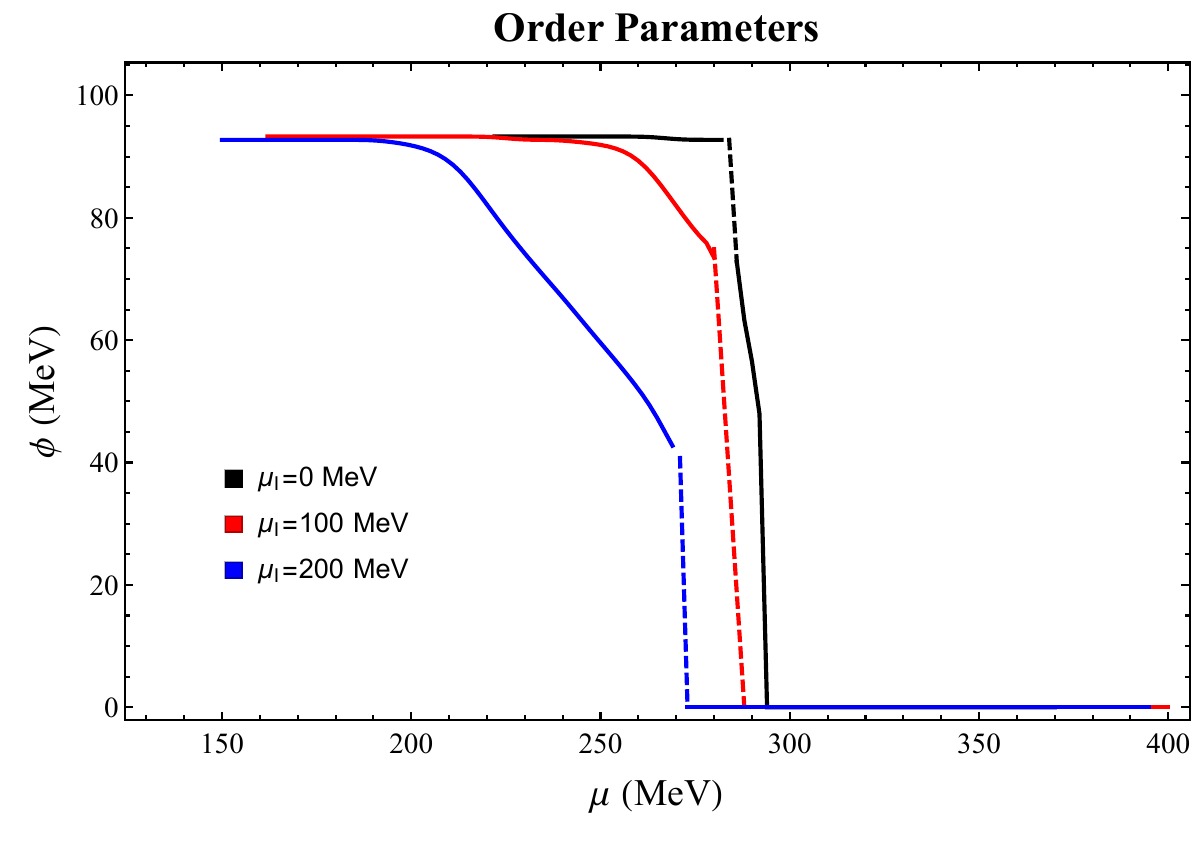}
\caption{\label{fig:5} 
 Chiral condensates as a function of quark chemical potential under different isospin chemical potentials, calculated at $T=10$ MeV, $g_{\omega}/m_{\omega}=g_{\rho}/m_{\rho}=0.006~[\mathrm{MeV}]^{-1}$. The different colored lines correspond to different isospin chemical potentials.
}
\end{figure}

\begin{figure}[!tbh]
\centering
\includegraphics[width=0.45\textwidth]{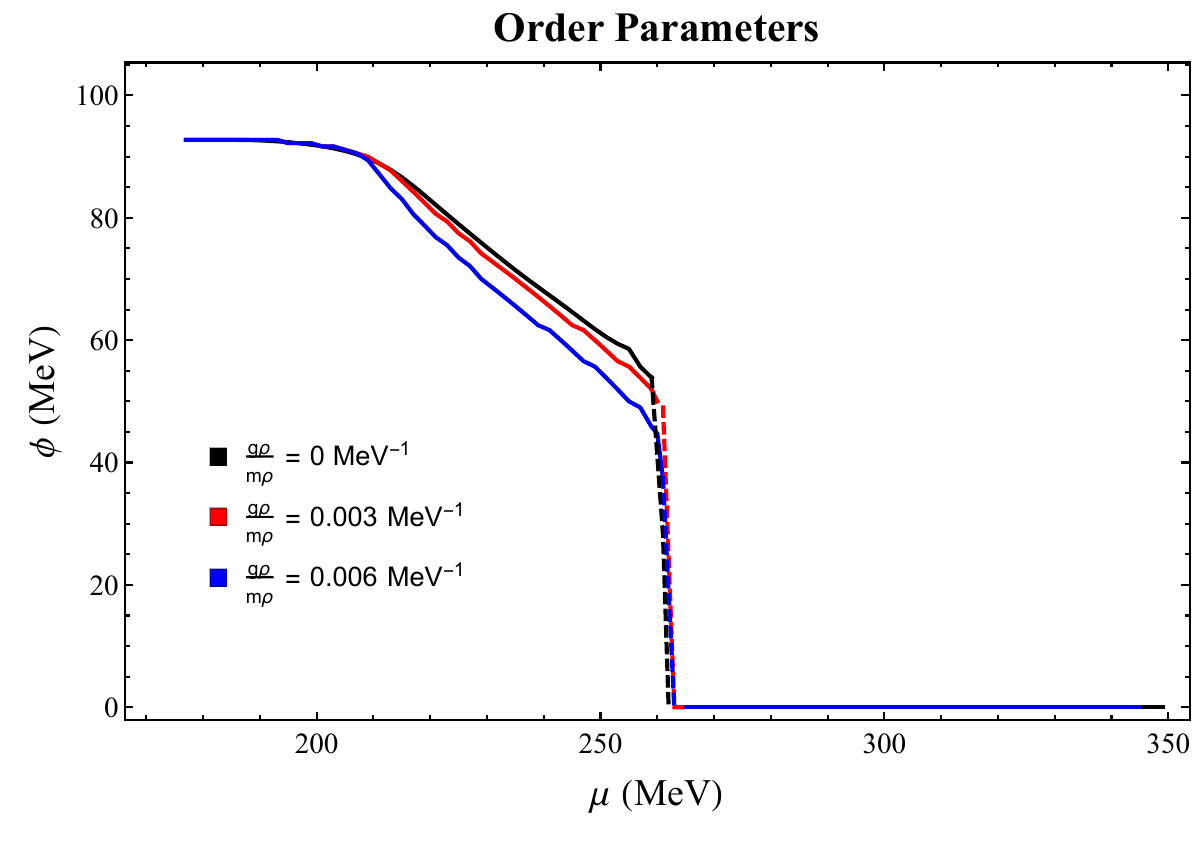}
\caption{\label{fig:6} Chiral condensates as a function of quark chemical potential calculated at $T=10$ MeV, $\mu_{I}=200$ MeV, $g_{\omega}/m_{\omega}=0~[\mathrm{MeV}]^{-1}$. The plot shows the variation for different $g_{\rho}/m_{\rho}$, different colored lines corresponding to different vector coupling strengths.}
\end{figure}

\begin{figure}[!tbh]
\centering
\includegraphics[width=0.45\textwidth]{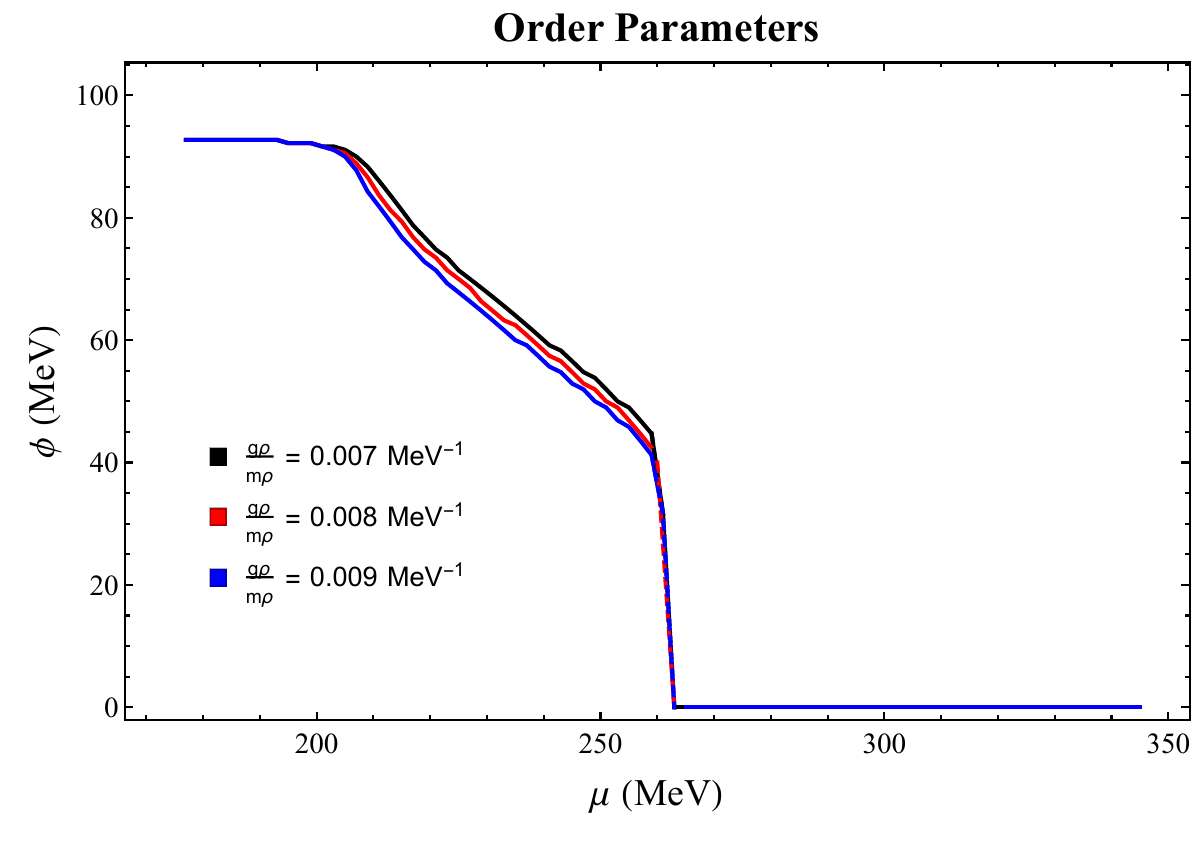}
\caption{\label{fig:7} Chiral condensates as a function of quark chemical potential calculated at $T=10$ MeV, $\mu_{I}=200$ MeV, $g_{\omega}/m_{\omega}=0~[\mathrm{MeV}]^{-1}$. The plot shows the variation for different $g_{\rho}/m_{\rho}$, different colored lines corresponding to different vector coupling strengths.}
\end{figure}

\begin{figure}[!tbh]
\centering
\includegraphics[width=0.45\textwidth]{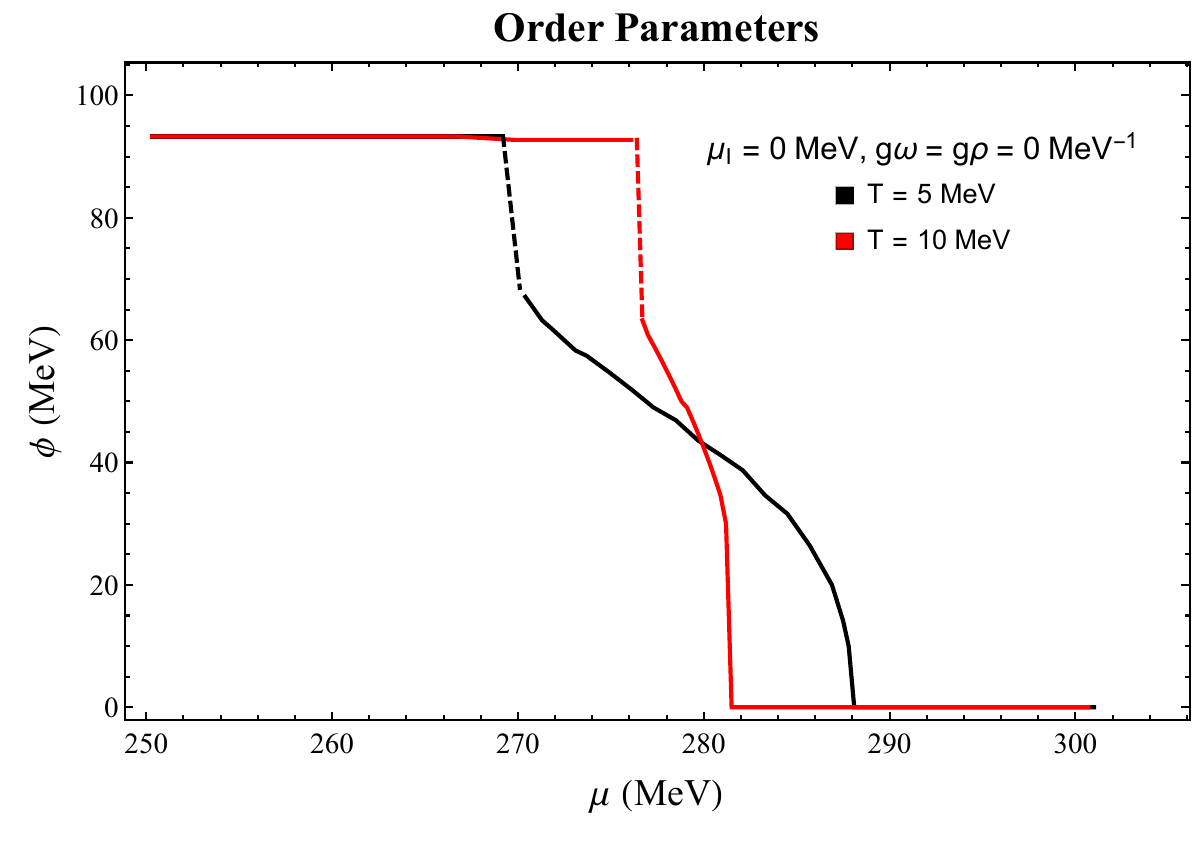}
\caption{\label{fig:9} Chiral condensates as a function of quark chemical potential, calculated when $\mu_{I}=0$ MeV and $g_{\omega}/m_{\omega}=g_{\rho}/m_{\rho}=0~[\mathrm{MeV}]^{-1}$. The black and red lines correspond to $T=5$ MeV and  $T=10$ MeV.}
\end{figure}

To investigate the nature of the chiral phase transition, we examine the behavior of the chiral condensate $\phi$ as a function of the quark chemical potential $\mu$ at low temperatures. 
Figs.~\ref{fig:4} and \ref{fig:5} display the condensate's evolution under different isospin chemical potentials. As $\mu_{I}$ increases, the transition occurs at lower values of $\mu$, resulting in a decrease in the initial values of the chiral condensate, and the initial values are lower at higher temperatures, which are closer to the second-order transition region, than at lower temperatures.
Figs.~\ref{fig:6} and \ref{fig:7} show the chiral condensate evolution for various $\rho$-meson coupling strengths, further highlighting the role of vector interactions. We find that with an increasing vector coupling, the transition types at smaller quark chemical potential positions continuously change from second-order to first-order. When the coupling constant is zero (including omega meson), but there exists a finite isospin chemical potential (depicted by the black line in Fig. \ref{fig:6}), it is fundamentally different from the matter that lacks both vector coupling and isospin chemical potential (depicted in Fig.~\ref{fig:9}). As we have observed in the results of the effective potential, these phenomena are evidently due to the modification of the vacuum near $\phi=0$ induced by the isospin chemical potential.


\begin{figure}[!tbh]
\centering
\includegraphics[width=0.45\textwidth]{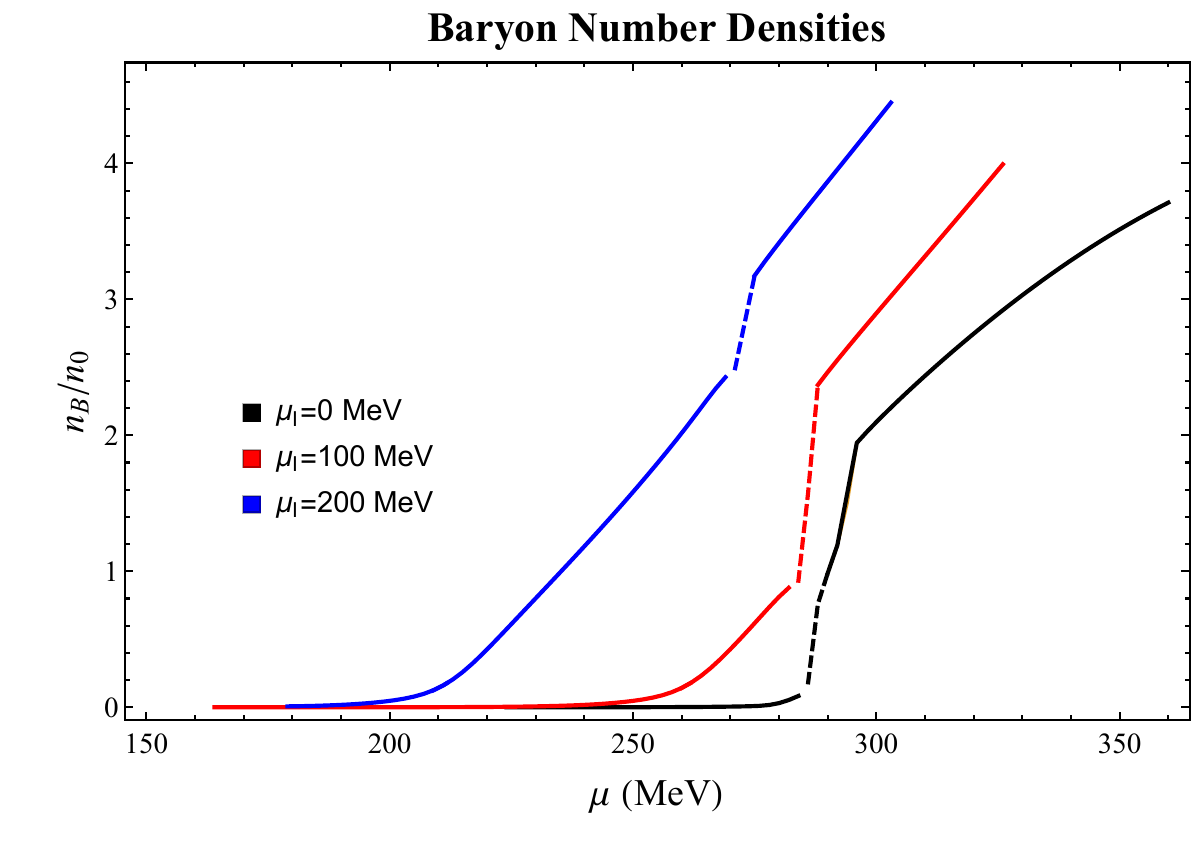}
\caption{The change of the baryon number densities with chemical potential when  $T = 10$ MeV. The black line corresponds to $\mu_{I}=0$ MeV, the blue line corresponds to $\mu_{I}=200$ MeV, the red line is in $\mu_{I}=300$ MeV, while the green line is in $\mu_{I}=400$ MeV. The baryon number density normalized by the nuclear saturation density $n_{0}=0.16 fm^{-1}$.}
\label{fig:8}
\end{figure}

\subsection{Baryon number density}
Fig.~\ref{fig:8} presents the baryon number density normalized by the nuclear saturation density $n_{0} = 0.16\,\text{fm}^{-3}$ as a function of $\mu$ at $T = 10$ MeV for different $\mu_{I}$ values.
% \ans{As $\mu_{I}$ increases, the transition eventually resembles a second-order behavior. Additionally, the onset of chiral phase transition shifts to lower quark $\mu$, hinting at the destabilizing influence of isospin asymmetry.}
At low $\mu_{I}$, a clear first-order phase transition is evident. As $\mu_{I}$ increases, the transition shows a tendency to approach a second-order transition.
% the transition weakens and eventually resembles a second-order behavior.
Additionally, the onset of chiral phase transition shifts to lower quark $\mu$, hinting at the destabilizing influence of isospin asymmetry.

\subsection{Phase diagram}
\begin{figure}[H]
\centering
\includegraphics[width=0.45\textwidth]{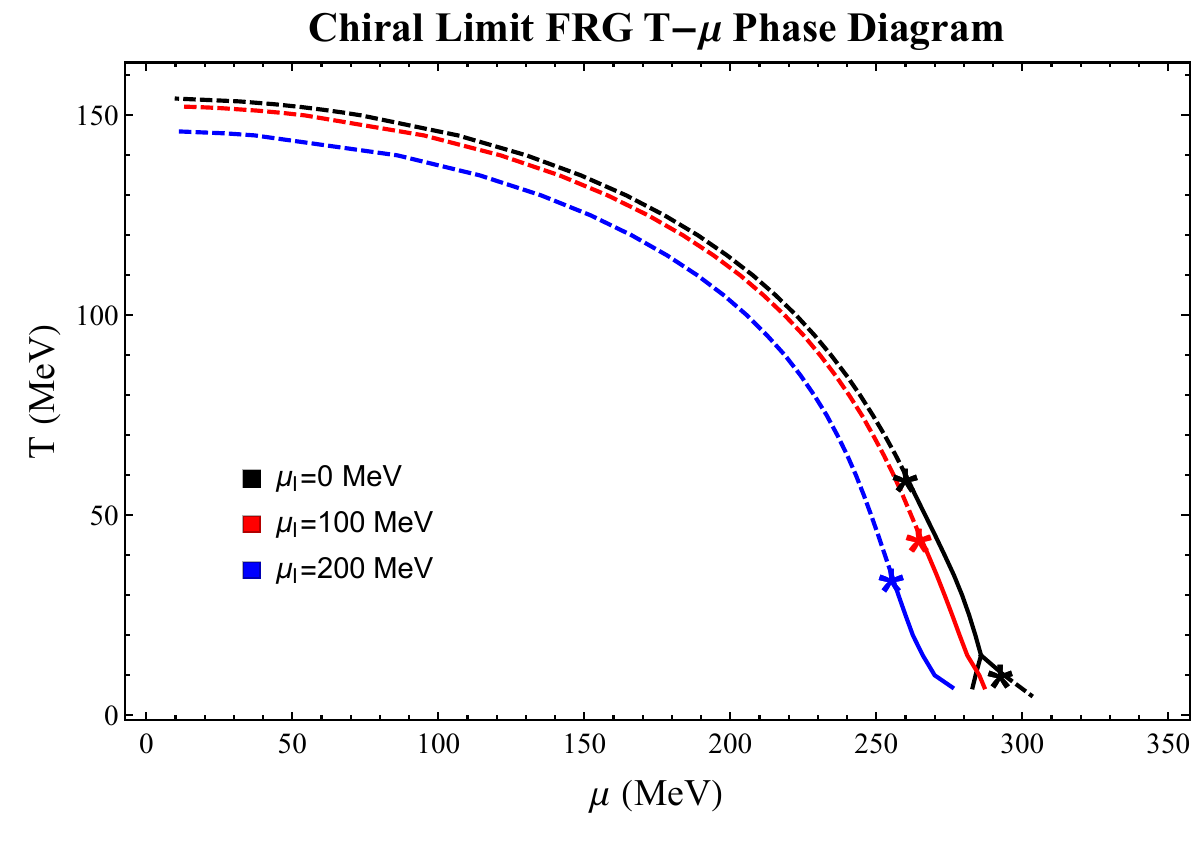}
\caption{\label{fig:10} The FRG ${T-\mu}$ phase diagram with different $\mu_{I}$. The solid lines represent the first-order phase transition, and the dashed lines represent the second-order phase transition. The stars show the TCPs. The parameters are set as: $f_{\pi} = 93$ MeV, $g_{s} = 3.2$, $\lambda = 8$, the ultraviolet cutoff $\Lambda_{FRG} = 500$ MeV, the coupling constants $g_{\omega}m^{-1}_{\omega} = g_{\rho} m^{-1}_{\rho} = 0.006 [\mathrm{MeV}]^{-1}$.}
\end{figure}

\begin{figure}[H]
\centering
\includegraphics[width=0.45\textwidth]{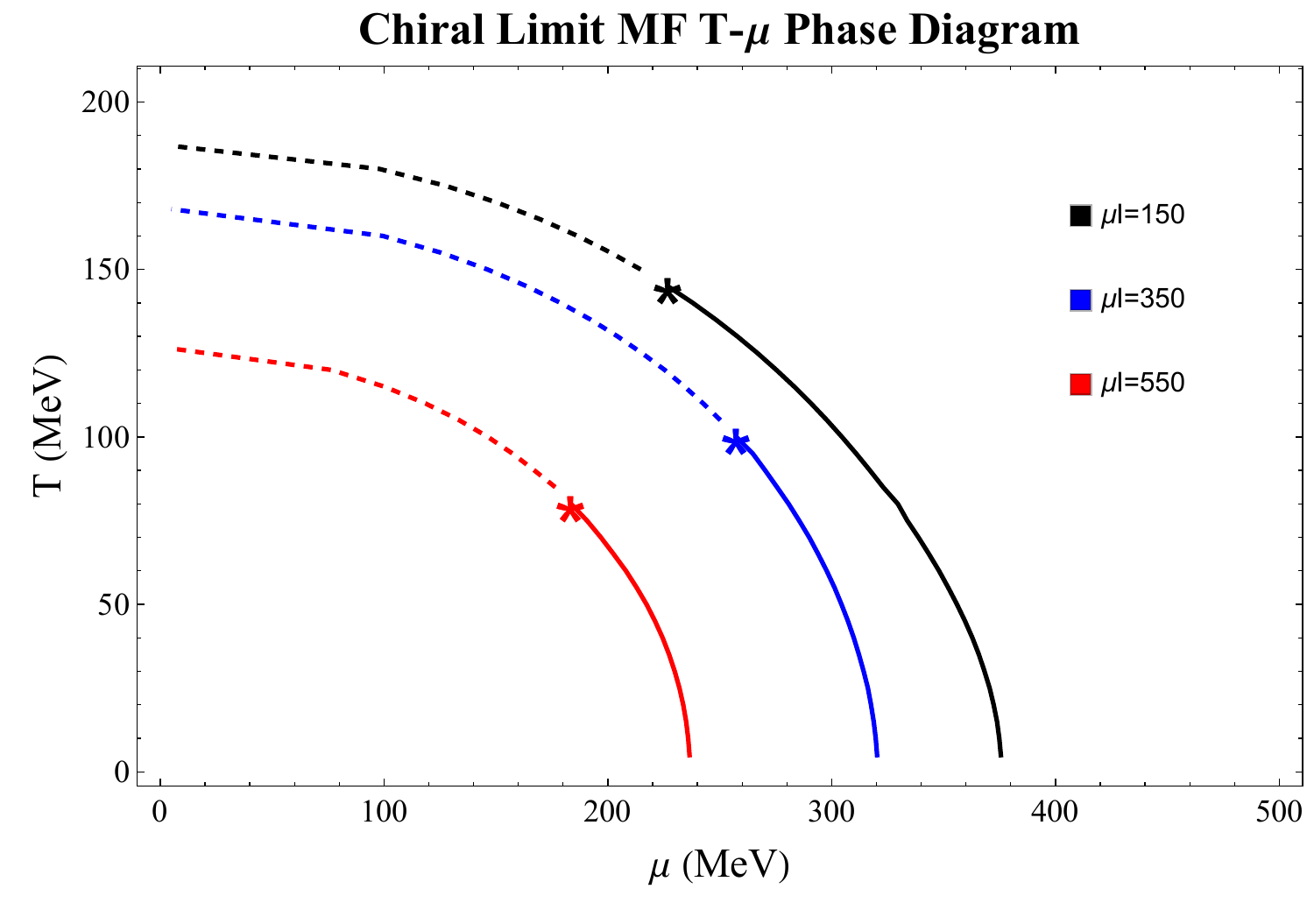}
\caption{\label{fig:11} The mean-field ${T-\mu}$ phase diagram including vacuum fluctuation ($\Lambda_{MF}=260$ MeV) for the two-flavor massless QCD with different isospin chemical potentials. Solid lines show the first-order phase transitions, and dashed lines show the second-order phase transition. Dashed lines show the critical points, and the stars show the position of the TCPs.}
\end{figure}

Figs.~\ref{fig:10} and \ref{fig:11} depict the phase diagrams in the $T-\mu$ plane, computed using the FRG and MF approaches for several values of the isospin chemical potential $\mu_{I}$. Solid lines denote first-order transitions, dashed lines represent second-order transitions, and stars indicate tricritical points (TCPs).
In the FRG case (Fig.~\ref{fig:10}), increasing $\mu_{I}$ shifts the phase boundary toward lower temperatures and chemical potentials, and also reduces the temperature of the TCP. This behavior is less pronounced in the MF results (Fig.~\ref{fig:11}), which lack fluctuation contributions and thus show a more conventional phase structure.

At $\mu_{I} = 0$, the $\rho$ meson does not contribute to the dynamics, and the only vector interaction arises from the $\omega$ meson.
In this limit, the FRG phase diagram exhibits the same " back-bending behavior"  identified in Ref.~\cite{Zhang:2017icm}, where the $\rho$ meson was included. This ``back-bending" refers to a reversal in the typical trend of the chiral transition line, where, instead of decreasing monotonically, the critical temperature briefly increases with increasing chemical potential. 

An intriguing feature in the FRG phase diagram we observe is that  the suppression of the ``back-bending" behavior at low temperatures when $\mu_{I}$ is increased. We also  find  that inclusion of the  $\rho $  vector meson in the model suppresses the ``back-bending" behavior of the phase boundary, in contrast to Ref.~\cite{Zhang:2017icm}, where $\rho$ was not considered.  This may due to the  physical assumptions and model scope limitations.
Underlying assumptions in the fermion sector may break down in certain regimes. Ref.~\cite{Tripolt:2017zgc} pointed out the presence of unphysical negative entropy in the low-temperature and high-density region of the phase diagram, attributing this behavior primarily to the quark-loop contribution in the flow equation.   Here the introduction of a finite isospin chemical potential leads to an evident improvement in the quark sector, effectively mitigating this issue. This  is  quite understandable physically .   For instanc at high density and low temperature, attractive pion and sigma exchange interactions could induce Cooper instabilities, leading to superconducting pairing gaps that coincide with regions of negative entropy. Finite isospin chemical potential and inclusion of  the  $\rho $  vector meson , which  suppress the Cooper-pair instabilities and thus tame the ``back-bending" behaviors, as shown explicitly in the phase diagram  Fig.~\ref{fig:10} .
%The results give us a glimpse into these complex phenomena.

Investigating the emergence of negative entropy and backbending phase boundaries in the low-temperature regime of the FRG phase diagram is a long-term  and intricate task.  Besides what we mentioned above here, the possible  reasons for the “backbending " behaviors include several  other  limitations in our current understanding.
The specific choice of regulator function could also affect low-energy predictions. An over-inclusion of quantum fluctuations during the evolution process, particularly at low energy scales where regulator cutoff effects become more pronounced. This interpretation aligns with recent theoretical work~\cite{Otto:2022jzl},which demonstrated that regulator dependence can significantly influence phase boundary determinations in the deep infrared regime.  In addition,  the observed "bending back" behavior may partially arise from truncation artifacts inherent in the derivative expansion. Development of improved truncation schemes may mitigate  these artifacts \cite{Ihssen:2023xlp}.

\section{SUMMARY}
\label{sec:4}
In this work, we have employed the functional renormalization group and mean field methods to investigate the two-flavor quark-meson model with omega and rho vector mesons, in the presence of a finite isospin chemical potential. Our goal was to explore how these vector couplings and isospin chemical potential influence the phase structure or the chiral phase diagram.

We found that vector mesons omega and rho have qualitatively distinct effects on the phase structure. While both shift the phase boundaries, the rho meson, in particular, plays a crucial role in modifying the critical endpoint and the nature of the phase transition.

Increasing the isospin chemical potential at a given vector coupling strength lowers the temperature and the baryon chemical potential  at which the chiral phase transition occurs. It also shift down the tricritical points on the phase diagram, effectively pushing the entire phase boundary toward lower $T$ and $\mu$ regions.

The functional renormalization group and mean field methods yield qualitatively different results. In particular, the FRG approach captures the typical  ``back-bending" feature in the low-temperature phase boundary. This "back-bending"  behaviors get suppressed  with the increasing $\mu_{I}$  and inclusion of the  vector meson $\rho$. Without the isospin chemical potential  and  vector meson $rho$ , the ``bending back" behavior is located at larger chemical potential, e.g., $280$ MeV to $300$ MeV. According to our results, the phase boundary shifts to a  smaller chemical potential after we introduce the vector meson $\rho$  and isospin chemical potential. In this sense, the ``bending back" behavior, of course, should be suppressed.

At the lower energy of   order  $( O(1) ) $  GeV energy scale of heavy-ion collisions, the baryon density and isospin density deviate from zero due to the stopping effect and the initial asymmetry in proton and neutron numbers of the colliding nuclei. Consequently, the role of the isospin chemical potential and vector coupling becomes non-negligible. At large baryon chemical potentials, these couplings introduce significant complexity into the system's physics. By incorporating the isospin chemical potential and vector coupling, we observed that the phase boundary shifts to a smaller chemical potential. This shift naturally suppresses the ``bending back" behavior. Physically, this improvement can be attributed to two key effects: (1) The suppression of Cooper pair instability due to the repulsive force mediated by the $\rho$ vector meson. (2) The Fermi surface imbalance induced by the finite isospin chemical potential. The extent to which the ``bending back" behavior is mitigated - or whether it vanishes entirely - depends on  the relative magnitudes of the vector coupling, isospin chemical potential, baryon chemical potential, and temperature. Further investigation is needed to quantify the precise relationship between these parameters and the resulting phase structure.

\acknowledgement{
\textbf{Acknowledgements} We thank Hai-cang Ren and Moran Jia for useful discussions. This work is supported in part by the National Key Research and
 Development Program of China under Contract
No. 2022YFA1604900. This work is also partly supported
by the National Natural Science Foundation of China
 (NSFC) under Grants No. 12435009, and No. 12275104. Hui Zhang acknowledges the financial support from the Guangdong Major Project of Basic and Applied Basic Research (Grant No. 2020B0301030008) and the National Natural Science Foundation of China (Grant No. 12047523, and 12105107).
}

%as required. Don't forget to give each section
%and subsection a unique label (see Sect.~\ref{sec:1}).
%
% % For one-column wide figures use
% \begin{figure}
% % Use the relevant command for your figure-insertion program
% % to insert the figure file.
% % For example, with the option graphics use
% \resizebox{0.75\textwidth}{!}{%
%   \includegraphics{leer.eps}
% }
% % If not, use
% %\vspace{5cm}       % Give the correct figure height in cm
% \caption{Please write your figure caption here}
% \label{fig:1}       % Give a unique label
% \end{figure}
% \begin{figure*}
% % Use the relevant command for your figure-insertion program
% % to insert the figure file. See example above.
% % If not, use
% \vspace*{5cm}       % Give the correct figure height in cm
% \caption{Please write your figure caption here}
% \label{fig:2}       % Give a unique label
% \end{figure*}
% \begin{table}
% \caption{Please write your table caption here}
% \label{tab:1}       % Give a unique label
% % For LaTeX tables use
% \begin{tabular}{lll}
% \hline\noalign{\smallskip}
% first & second & third  \\
% \noalign{\smallskip}\hline\noalign{\smallskip}
% number & number & number \\
% number & number & number \\
% \noalign{\smallskip}\hline
% \end{tabular}
% % Or use
% \vspace*{5cm}  % with the correct table height
% \end{table}
%
% BibTeX users please use
\bibliographystyle{utphys}
\bibliography{references}

\providecommand{\href}[2]{#2}\begingroup\raggedright\begin{thebibliography}{10}

\bibitem{Braun-Munzinger:2008szb}
P.~Braun-Munzinger and J.~Wambach, ``{The Phase Diagram of Strongly-Interacting Matter}'', \href{http://dx.doi.org/10.1103/RevModPhys.81.1031}{{\em Rev. Mod. Phys.} {\bfseries 81} (2009) }, \href{http://arxiv.org/abs/0801.4256}{{\ttfamily arXiv:0801.4256 [hep-ph]}}.

\bibitem{Rennecke:2015eba}
F.~Rennecke, ``{Vacuum structure of vector mesons in QCD}'', \href{http://dx.doi.org/10.1103/PhysRevD.92.076012}{{\em Phys. Rev. D} {\bfseries 92} no.~7, (2015) }, \href{http://arxiv.org/abs/1504.03585}{{\ttfamily arXiv:1504.03585 [hep-ph]}}.

\bibitem{Cabibbo:1975ig}
N.~Cabibbo and G.~Parisi, ``{Exponential Hadronic Spectrum and Quark Liberation}'', \href{http://dx.doi.org/10.1016/0370-2693(75)90158-6}{{\em Phys. Lett. B} {\bfseries 59} (1975) }.

\bibitem{deForcrand:2009zkb}
P.~de~Forcrand, ``{Simulating QCD at finite density}'', \href{http://dx.doi.org/10.22323/1.091.0010}{{\em PoS} {\bfseries LAT2009} (2009) }, \href{http://arxiv.org/abs/1005.0539}{{\ttfamily arXiv:1005.0539 [hep-lat]}}.

\bibitem{STAR:2011hyh}
{\bfseries STAR} Collaboration, L.~Adamczyk {\em et~al.}, ``{Directed Flow of Identified Particles in Au + Au Collisions at $\sqrt{s{_NN}} = 200$ GeV at RHIC}'', \href{http://dx.doi.org/10.1103/PhysRevLett.108.202301}{{\em Phys. Rev. Lett.} {\bfseries 108} (2012) }, \href{http://arxiv.org/abs/1112.3930}{{\ttfamily arXiv:1112.3930 [nucl-ex]}}.

\bibitem{STAR:2014clz}
{\bfseries STAR} Collaboration, L.~Adamczyk {\em et~al.}, ``{Beam-Energy Dependence of the Directed Flow of Protons, Antiprotons, and Pions in Au+Au Collisions}'', \href{http://dx.doi.org/10.1103/PhysRevLett.112.162301}{{\em Phys. Rev. Lett.} {\bfseries 112} no.~16, (2014) }, \href{http://arxiv.org/abs/1401.3043}{{\ttfamily arXiv:1401.3043 [nucl-ex]}}.

\bibitem{STAR:2017tfy}
{\bfseries STAR} Collaboration, L.~Adamczyk {\em et~al.}, ``{Collision Energy Dependence of Moments of Net-Kaon Multiplicity Distributions at RHIC}'', \href{http://dx.doi.org/10.1016/j.physletb.2018.07.066}{{\em Phys. Lett. B} {\bfseries 785} (2018) }, \href{http://arxiv.org/abs/1709.00773}{{\ttfamily arXiv:1709.00773 [nucl-ex]}}.

\bibitem{STAR:2017sal}
{\bfseries STAR} Collaboration, L.~Adamczyk {\em et~al.}, ``{Bulk Properties of the Medium Produced in Relativistic Heavy-Ion Collisions from the Beam Energy Scan Program}'', \href{http://dx.doi.org/10.1103/PhysRevC.96.044904}{{\em Phys. Rev. C} {\bfseries 96} no.~4, (2017) }, \href{http://arxiv.org/abs/1701.07065}{{\ttfamily arXiv:1701.07065 [nucl-ex]}}.

\bibitem{NA61SHINE:2021wba}
{\bfseries NA61/SHINE} Collaboration, A.~Acharya {\em et~al.}, ``{$K^{*}(892)^0$ meson production in inelastic p+p interactions at $40$ and $80$~GeV/$c$ beam momenta measured by NA61/SHINE at the CERN SPS}'', \href{http://dx.doi.org/10.1140/epjc/s10052-022-10281-5}{{\em Eur. Phys. J. C} {\bfseries 82} no.~4, (2022) }, \href{http://arxiv.org/abs/2112.09506}{{\ttfamily arXiv:2112.09506 [nucl-ex]}}.

\bibitem{Grebieszkow:2017gqx}
{\bfseries NA61/SHINE} Collaboration, K.~Grebieszkow, ``{New results on fluctuations and correlations from the NA61/SHINE experiment at the CERN SPS}'', \href{http://dx.doi.org/10.22323/1.314.0167}{{\em PoS} {\bfseries EPS-HEP2017} (2017) }, \href{http://arxiv.org/abs/1709.10397}{{\ttfamily arXiv:1709.10397 [nucl-ex]}}.

\bibitem{blaschke2016topical}
D.~Blaschke, J.~Aichelin, E.~Bratkovskaya, V.~Friese, M.~Gazdzicki, J.~Randrup, O.~Rogachevsky, O.~Teryaev, and V.~Toneev, ``{Topical issue on Exploring Strongly Interacting Matter at High Densities - NICA White Paper}'', \href{http://dx.doi.org/10.1140/epja/i2016-16267-x}{{\em Eur. Phys. J. A} {\bfseries 52} no.~8, (2016) }.

\bibitem{CBM:2016kpk}
{\bfseries CBM} Collaboration, T.~Ablyazimov {\em et~al.}, ``{Challenges in QCD matter physics --The scientific programme of the Compressed Baryonic Matter experiment at FAIR}'', \href{http://dx.doi.org/10.1140/epja/i2017-12248-y}{{\em Eur. Phys. J. A} {\bfseries 53} no.~3, (2017) }, \href{http://arxiv.org/abs/1607.01487}{{\ttfamily arXiv:1607.01487 [nucl-ex]}}.

\bibitem{SAKO20141158}
H.~Sako {\em et~al.}, ``{Towards the heavy-ion program at J-PARC}'', \href{http://dx.doi.org/10.1016/j.nuclphysa.2014.08.065}{{\em Nucl. Phys. A} {\bfseries 931} (2014) }.

\bibitem{Nikolov:1996jj}
E.~N. Nikolov, W.~Broniowski, C.~V. Christov, G.~Ripka, and K.~Goeke, ``{Meson loops in the Nambu-Jona-Lasinio model}'', \href{http://dx.doi.org/10.1016/0375-9474(96)00231-X}{{\em Nucl. Phys. A} {\bfseries 608} (1996) }, \href{http://arxiv.org/abs/hep-ph/9602274}{{\ttfamily arXiv:hep-ph/9602274}}.

\bibitem{Nemoto:1999qf}
Y.~Nemoto, K.~Naito, and M.~Oka, ``{Effective potential of O(N) linear sigma model at finite temperature}'', \href{http://dx.doi.org/10.1007/s100500070042}{{\em Eur. Phys. J. A} {\bfseries 9} (2000) }, \href{http://arxiv.org/abs/hep-ph/9911431}{{\ttfamily arXiv:hep-ph/9911431}}.

\bibitem{Oertel:2000jp}
M.~Oertel, M.~Buballa, and J.~Wambach, ``{Meson loop effects in the NJL model at zero and nonzero temperature}'', \href{http://dx.doi.org/10.1134/1.1368226}{{\em Phys. Atom. Nucl.} {\bfseries 64} (2001) }, \href{http://arxiv.org/abs/hep-ph/0008131}{{\ttfamily arXiv:hep-ph/0008131}}.

\bibitem{Baacke:2003dk}
J.~Baacke and S.~Michalski, ``{O(N) linear sigma model beyond the Hartree approximation at finite temperature}'', in {\em {6th Workshop on Quantum Field Theory under the Influence of External Conditions (QFEXT03)}}, pp.~282--287.
\newblock 12, 2003.
\newblock \href{http://arxiv.org/abs/hep-ph/0312031}{{\ttfamily arXiv:hep-ph/0312031}}.

\bibitem{Andersen:2008qk}
J.~O. Andersen and T.~Brauner, ``{Linear sigma model at finite density in the 1/N expansion to next-to-leading order}'', \href{http://dx.doi.org/10.1103/PhysRevD.78.014030}{{\em Phys. Rev. D} {\bfseries 78} (2008) }, \href{http://arxiv.org/abs/0804.4604}{{\ttfamily arXiv:0804.4604 [hep-ph]}}.

\bibitem{Muller:2010am}
D.~M{\"u}ller, M.~Buballa, and J.~Wambach, ``{The Quark Propagator in the NJL Model in a self-consistent 1/Nc Expansion}'', \href{http://dx.doi.org/10.1103/PhysRevD.81.094022}{{\em Phys. Rev. D} {\bfseries 81} (2010) }, \href{http://arxiv.org/abs/1002.4252}{{\ttfamily arXiv:1002.4252 [hep-ph]}}.

\bibitem{Yamazaki:2012ux}
K.~Yamazaki and T.~Matsui, ``{Quark-Hadron Phase Transition in the PNJL model for interacting quarks}'', \href{http://dx.doi.org/10.1016/j.nuclphysa.2013.05.018}{{\em Nucl. Phys. A} {\bfseries 913} (2013) }, \href{http://arxiv.org/abs/1212.6165}{{\ttfamily arXiv:1212.6165 [hep-ph]}}.

\bibitem{Zacchi:2017ahv}
A.~Zacchi and J.~Schaffner-Bielich, ``{Effects of Renormalizing the chiral SU(2) Quark-Meson-Model}'', \href{http://dx.doi.org/10.1103/PhysRevD.97.074011}{{\em Phys. Rev. D} {\bfseries 97} no.~7, (2018) }, \href{http://arxiv.org/abs/1712.01629}{{\ttfamily arXiv:1712.01629 [hep-ph]}}.

\bibitem{CamaraPereira:2020ipu}
R.~C\^amara~Pereira and P.~Costa, ``{One-meson-loop NJL model: Effect of collective and noncollective excitations on the quark condensate at finite temperature}'', \href{http://dx.doi.org/10.1103/PhysRevD.101.054025}{{\em Phys. Rev. D} {\bfseries 101} no.~5, (2020) }, \href{http://arxiv.org/abs/2003.08430}{{\ttfamily arXiv:2003.08430 [hep-ph]}}.

\bibitem{Dupuis:2020fhh}
N.~Dupuis, L.~Canet, A.~Eichhorn, W.~Metzner, J.~M. Pawlowski, M.~Tissier, and N.~Wschebor, ``{The nonperturbative functional renormalization group and its applications}'', \href{http://dx.doi.org/10.1016/j.physrep.2021.01.001}{{\em Phys. Rept.} {\bfseries 910} (2021) }, \href{http://arxiv.org/abs/2006.04853}{{\ttfamily arXiv:2006.04853 [cond-mat.stat-mech]}}.

\bibitem{Fukushima:2012xw}
K.~Fukushima and J.~M. Pawlowski, ``{Magnetic catalysis in hot and dense quark matter and quantum fluctuations}'', \href{http://dx.doi.org/10.1103/PhysRevD.86.076013}{{\em Phys. Rev. D} {\bfseries 86} (2012) }, \href{http://arxiv.org/abs/1203.4330}{{\ttfamily arXiv:1203.4330 [hep-ph]}}.

\bibitem{Braun:2011pp}
J.~Braun, ``{Fermion Interactions and Universal Behavior in Strongly Interacting Theories}'', \href{http://dx.doi.org/10.1088/0954-3899/39/3/033001}{{\em J. Phys. G} {\bfseries 39} (2012) }, \href{http://arxiv.org/abs/1108.4449}{{\ttfamily arXiv:1108.4449 [hep-ph]}}.

\bibitem{Aoki:2014ola}
K.-I. Aoki, S.-I. Kumamoto, and D.~Sato, ``{Weak solution of the non-perturbative renormalization group equation to describe dynamical chiral symmetry breaking}'', \href{http://dx.doi.org/10.1093/ptep/ptu039}{{\em PTEP} {\bfseries 2014} no.~4, (2014) }, \href{http://arxiv.org/abs/1403.0174}{{\ttfamily arXiv:1403.0174 [hep-th]}}.

\bibitem{Aoki:2015hsa}
K.-I. Aoki and M.~Yamada, ``{The RG flow of Nambu\textendash{}Jona-Lasinio model at finite temperature and density}'', \href{http://dx.doi.org/10.1142/S0217751X15501808}{{\em Int. J. Mod. Phys. A} {\bfseries 30} no.~27, (2015) }, \href{http://arxiv.org/abs/1504.00749}{{\ttfamily arXiv:1504.00749 [hep-ph]}}.

\bibitem{Schaefer:2004en}
B.-J. Schaefer and J.~Wambach, ``{The Phase diagram of the quark meson model}'', \href{http://dx.doi.org/10.1016/j.nuclphysa.2005.04.012}{{\em Nucl. Phys. A} {\bfseries 757} (2005) }, \href{http://arxiv.org/abs/nucl-th/0403039}{{\ttfamily arXiv:nucl-th/0403039}}.

\bibitem{Herbst:2013ail}
T.~K. Herbst, J.~M. Pawlowski, and B.-J. Schaefer, ``{Phase structure and thermodynamics of QCD}'', \href{http://dx.doi.org/10.1103/PhysRevD.88.014007}{{\em Phys. Rev. D} {\bfseries 88} no.~1, (2013) }, \href{http://arxiv.org/abs/1302.1426}{{\ttfamily arXiv:1302.1426 [hep-ph]}}.

\bibitem{Fu:2015naa}
W.-j. Fu and J.~M. Pawlowski, ``{Relevance of matter and glue dynamics for baryon number fluctuations}'', \href{http://dx.doi.org/10.1103/PhysRevD.92.116006}{{\em Phys. Rev. D} {\bfseries 92} no.~11, (2015) }, \href{http://arxiv.org/abs/1508.06504}{{\ttfamily arXiv:1508.06504 [hep-ph]}}.

\bibitem{Herbst:2013ufa}
T.~K. Herbst, M.~Mitter, J.~M. Pawlowski, B.-J. Schaefer, and R.~Stiele, ``{Thermodynamics of QCD at vanishing density}'', \href{http://dx.doi.org/10.1016/j.physletb.2014.02.045}{{\em Phys. Lett. B} {\bfseries 731} (2014) }, \href{http://arxiv.org/abs/1308.3621}{{\ttfamily arXiv:1308.3621 [hep-ph]}}.

\bibitem{Tripolt:2013jra}
R.-A. Tripolt, N.~Strodthoff, L.~von Smekal, and J.~Wambach, ``{Spectral Functions for the Quark-Meson Model Phase Diagram from the Functional Renormalization Group}'', \href{http://dx.doi.org/10.1103/PhysRevD.89.034010}{{\em Phys. Rev. D} {\bfseries 89} no.~3, (2014) }, \href{http://arxiv.org/abs/1311.0630}{{\ttfamily arXiv:1311.0630 [hep-ph]}}.

\bibitem{Jung:2016yxl}
C.~Jung, F.~Rennecke, R.-A. Tripolt, L.~von Smekal, and J.~Wambach, ``{In-Medium Spectral Functions of Vector- and Axial-Vector Mesons from the Functional Renormalization Group}'', \href{http://dx.doi.org/10.1103/PhysRevD.95.036020}{{\em Phys. Rev. D} {\bfseries 95} no.~3, (2017) }, \href{http://arxiv.org/abs/1610.08754}{{\ttfamily arXiv:1610.08754 [hep-ph]}}.

\bibitem{Andersen:2013swa}
J.~O. Andersen, W.~R. Naylor, and A.~Tranberg, ``{Chiral and deconfinement transitions in a magnetic background using the functional renormalization group with the Polyakov loop}'', \href{http://dx.doi.org/10.1007/JHEP04(2014)187}{{\em JHEP} {\bfseries 04} (2014) }, \href{http://arxiv.org/abs/1311.2093}{{\ttfamily arXiv:1311.2093 [hep-ph]}}.

\bibitem{Schaefer:2006sr}
B.-J. Schaefer and J.~Wambach, ``{Renormalization group approach towards the QCD phase diagram}'', \href{http://dx.doi.org/10.1134/S1063779608070083}{{\em Phys. Part. Nucl.} {\bfseries 39} (2008) }, \href{http://arxiv.org/abs/hep-ph/0611191}{{\ttfamily arXiv:hep-ph/0611191}}.

\bibitem{Herbst:2010rf}
T.~K. Herbst, J.~M. Pawlowski, and B.-J. Schaefer, ``{The phase structure of the Polyakov\textendash{}quark\textendash{}meson model beyond mean field}'', \href{http://dx.doi.org/10.1016/j.physletb.2010.12.003}{{\em Phys. Lett. B} {\bfseries 696} (2011) }, \href{http://arxiv.org/abs/1008.0081}{{\ttfamily arXiv:1008.0081 [hep-ph]}}.

\bibitem{Eser:2015pka}
J.~Eser, M.~Grahl, and D.~H. Rischke, ``{Functional Renormalization Group Study of the Chiral Phase Transition Including Vector and Axial-vector Mesons}'', \href{http://dx.doi.org/10.1103/PhysRevD.92.096008}{{\em Phys. Rev. D} {\bfseries 92} no.~9, (2015) }, \href{http://arxiv.org/abs/1508.06928}{{\ttfamily arXiv:1508.06928 [hep-ph]}}.

\bibitem{Drews:2014wba}
M.~Drews and W.~Weise, ``{Functional renormalization group approach to neutron matter}'', \href{http://dx.doi.org/10.1016/j.physletb.2014.09.051}{{\em Phys. Lett. B} {\bfseries 738} (2014) }, \href{http://arxiv.org/abs/1404.0882}{{\ttfamily arXiv:1404.0882 [nucl-th]}}.

\bibitem{Drews:2014spa}
M.~Drews and W.~Weise, ``{From asymmetric nuclear matter to neutron stars: a functional renormalization group study}'', \href{http://dx.doi.org/10.1103/PhysRevC.91.035802}{{\em Phys. Rev. C} {\bfseries 91} no.~3, (2015) }, \href{http://arxiv.org/abs/1412.7655}{{\ttfamily arXiv:1412.7655 [nucl-th]}}.

\bibitem{Drews:2013hha}
M.~Drews, T.~Hell, B.~Klein, and W.~Weise, ``{Thermodynamic phases and mesonic fluctuations in a chiral nucleon-meson model}'', \href{http://dx.doi.org/10.1103/PhysRevD.88.096011}{{\em Phys. Rev. D} {\bfseries 88} no.~9, (2013) }, \href{http://arxiv.org/abs/1308.5596}{{\ttfamily arXiv:1308.5596 [hep-ph]}}.

\bibitem{Fukushima:2008wg}
K.~Fukushima, ``{Phase diagrams in the three-flavor Nambu-Jona-Lasinio model with the Polyakov loop}'', \href{http://dx.doi.org/10.1103/PhysRevD.77.114028}{{\em Phys. Rev. D} {\bfseries 77} (2008) }, \href{http://arxiv.org/abs/0803.3318}{{\ttfamily arXiv:0803.3318 [hep-ph]}}. [Erratum: Phys.Rev.D 78, 039902 (2008)].

\bibitem{Bratovic:2012qs}
N.~M. Bratovic, T.~Hatsuda, and W.~Weise, ``{Role of Vector Interaction and Axial Anomaly in the PNJL Modeling of the QCD Phase Diagram}'', \href{http://dx.doi.org/10.1016/j.physletb.2013.01.003}{{\em Phys. Lett. B} {\bfseries 719} (2013) }, \href{http://arxiv.org/abs/1204.3788}{{\ttfamily arXiv:1204.3788 [hep-ph]}}.

\bibitem{Lourenco:2012yv}
O.~Lourenco, M.~Dutra, T.~Frederico, A.~Delfino, and M.~Malheiro, ``{Vector interaction strength in Polyakov-Nambu-Jona-Lasinio models from hadron-quark phase diagrams}'', \href{http://dx.doi.org/10.1103/PhysRevD.85.097504}{{\em Phys. Rev. D} {\bfseries 85} (2012) }, \href{http://arxiv.org/abs/1204.6357}{{\ttfamily arXiv:1204.6357 [nucl-th]}}.

\bibitem{Lu:2015naa}
Y.~Lu, Y.-L. Du, Z.-F. Cui, and H.-S. Zong, ``{Critical behaviors near the (tri-)critical end point of QCD within the NJL model}'', \href{http://dx.doi.org/10.1140/epjc/s10052-015-3720-2}{{\em Eur. Phys. J. C} {\bfseries 75} no.~10, (2015) }, \href{http://arxiv.org/abs/1508.00651}{{\ttfamily arXiv:1508.00651 [hep-ph]}}.

\bibitem{Adhikari:2017ydi}
P.~Adhikari, J.~O. Andersen, and P.~Kneschke, ``{Inhomogeneous chiral condensate in the quark-meson model}'', \href{http://dx.doi.org/10.1103/PhysRevD.96.016013}{{\em Phys. Rev. D} {\bfseries 96} no.~1, (2017) }, \href{http://arxiv.org/abs/1702.01324}{{\ttfamily arXiv:1702.01324 [hep-ph]}}. [Erratum: Phys.Rev.D 98, 099902 (2018)].

\bibitem{Floerchinger:2012xd}
S.~Floerchinger and C.~Wetterich, ``{Chemical freeze-out in heavy ion collisions at large baryon densities}'', \href{http://dx.doi.org/10.1016/j.nuclphysa.2012.07.009}{{\em Nucl. Phys. A} {\bfseries 890-891} (2012) }, \href{http://arxiv.org/abs/1202.1671}{{\ttfamily arXiv:1202.1671 [nucl-th]}}.

\bibitem{Skokov:2010sf}
V.~Skokov, B.~Friman, E.~Nakano, K.~Redlich, and B.~J. Schaefer, ``{Vacuum fluctuations and the thermodynamics of chiral models}'', \href{http://dx.doi.org/10.1103/PhysRevD.82.034029}{{\em Phys. Rev. D} {\bfseries 82} (2010) }, \href{http://arxiv.org/abs/1005.3166}{{\ttfamily arXiv:1005.3166 [hep-ph]}}.

\bibitem{Scavenius:2000qd}
O.~Scavenius, A.~Mocsy, I.~N. Mishustin, and D.~H. Rischke, ``{Chiral phase transition within effective models with constituent quarks}'', \href{http://dx.doi.org/10.1103/PhysRevC.64.045202}{{\em Phys. Rev. C} {\bfseries 64} (2001) }, \href{http://arxiv.org/abs/nucl-th/0007030}{{\ttfamily arXiv:nucl-th/0007030}}.

\bibitem{Zhang:2017icm}
H.~Zhang, D.~Hou, T.~Kojo, and B.~Qin, ``{Functional renormalization group study of the quark-meson model with $\omega$ meson}'', \href{http://dx.doi.org/10.1103/PhysRevD.96.114029}{{\em Phys. Rev. D} {\bfseries 96} no.~11, (2017) }, \href{http://arxiv.org/abs/1709.05654}{{\ttfamily arXiv:1709.05654 [hep-ph]}}.

\bibitem{Berges:2000ew}
J.~Berges, N.~Tetradis, and C.~Wetterich, ``{Nonperturbative renormalization flow in quantum field theory and statistical physics}'', \href{http://dx.doi.org/10.1016/S0370-1573(01)00098-9}{{\em Phys. Rept.} {\bfseries 363} (2002) }, \href{http://arxiv.org/abs/hep-ph/0005122}{{\ttfamily arXiv:hep-ph/0005122}}.

\bibitem{tripolt2014spectral}
R.-A. Tripolt, N.~Strodthoff, L.~von Smekal, and J.~Wambach, ``Spectral functions for the quark-meson model phase diagram from the functional renormalization group'', {\em Physical Review D} {\bfseries 89} no.~3, (2014) .

\bibitem{Tripolt:2017zgc}
R.-A. Tripolt, B.-J. Schaefer, L.~von Smekal, and J.~Wambach, ``{Low-temperature behavior of the quark-meson model}'', \href{http://dx.doi.org/10.1103/PhysRevD.97.034022}{{\em Phys. Rev. D} {\bfseries 97} no.~3, (2018) }, \href{http://arxiv.org/abs/1709.05991}{{\ttfamily arXiv:1709.05991 [hep-ph]}}.

\bibitem{Strodthoff:2013cua}
N.~Strodthoff and L.~von Smekal, ``{Polyakov-Quark-Meson-Diquark Model for two-color QCD}'', \href{http://dx.doi.org/10.1016/j.physletb.2014.03.008}{{\em Phys. Lett. B} {\bfseries 731} (2014) }, \href{http://arxiv.org/abs/1306.2897}{{\ttfamily arXiv:1306.2897 [hep-ph]}}.

\bibitem{Litim:2001up}
D.~F. Litim, ``{Optimized renormalization group flows}'', \href{http://dx.doi.org/10.1103/PhysRevD.64.105007}{{\em Phys. Rev. D} {\bfseries 64} (2001) }, \href{http://arxiv.org/abs/hep-th/0103195}{{\ttfamily arXiv:hep-th/0103195}}.

\bibitem{Otto:2022jzl}
K.~Otto, C.~Busch, and B.-J. Schaefer, ``{Regulator scheme dependence of the chiral phase transition at high densities}'', \href{http://dx.doi.org/10.1103/PhysRevD.106.094018}{{\em Phys. Rev. D} {\bfseries 106} no.~9, (2022) }, \href{http://arxiv.org/abs/2206.13067}{{\ttfamily arXiv:2206.13067 [hep-ph]}}.

\bibitem{Ihssen:2023xlp}
F.~Ihssen, J.~M. Pawlowski, F.~R. Sattler, and N.~Wink, ``{Toward quantitative precision for QCD at large densities}'', \href{http://dx.doi.org/10.1103/PhysRevD.111.036030}{{\em Phys. Rev. D} {\bfseries 111} no.~3, (2025) }, \href{http://arxiv.org/abs/2309.07335}{{\ttfamily arXiv:2309.07335 [hep-th]}}.

\end{thebibliography}\endgroup
% \begin{thebibliography}{}
% %
% % and use \bibitem to create references.
% % Format for Journal Reference
% \textbf{Volume}, (year) page numbers.
% Author2, \textit{Book title} (Publisher, place year) page numbers
% \end{thebibliography}
\end{document}